\documentclass[10pt,logo,copyright]{nvidiatechreport}
\usepackage[authoryear,round]{natbib}

\usepackage[utf8]{inputenc} 
\usepackage[T1]{fontenc}    

\usepackage{parskip}        
\usepackage{url}            
\usepackage{booktabs}       
\usepackage{amsfonts}       
\usepackage{nicefrac}       
\usepackage{microtype}      
\usepackage{xcolor}         
\usepackage[dvipsnames]{xcolor} 
\usepackage{graphicx}
\usepackage{animate}        
\usepackage{subcaption}
\usepackage{tabularx}
\usepackage{makecell}
\usepackage{adjustbox}
\usepackage{setspace}
\newcolumntype{M}[1]{>{\centering\arraybackslash}m{#1}}
\usepackage{float}
\usepackage{tikz}
\usetikzlibrary{positioning,shapes,arrows}
\usepackage{amsmath,amsfonts,bm, bbm,leftindex}
\usepackage{multirow}
\usepackage{comment}
\usepackage{gensymb}
\usepackage{lipsum}
\usetikzlibrary{arrows.meta, positioning, fit}
\usepackage[para]{threeparttable}
\usepackage{tikz}
\usetikzlibrary{tikzmark}

\usepackage[most]{tcolorbox}
\usepackage{fancyvrb}
\usepackage{fvextra}
\usepackage{dashrule}

\newcommand{\nvGreen}[1]{\textcolor{nvidiagreen}{#1}}

\def\eqref#1{equation~\ref{#1}}

\def\1{\bm{1}}

\def\vb{{\bm{b}}}
\def\vc{{\bm{c}}}
\def\vd{{\bm{d}}}
\def\ve{{\bm{e}}}

\def\vh{{\bm{h}}}
\def\vi{{\bm{i}}}

\def\vt{{\bm{t}}}

\def\vv{{\bm{v}}}
\def\vw{{\bm{w}}}
\def\vx{{\bm{x}}}

\def\mA{{\bm{A}}}

\def\mD{{\bm{D}}}
\def\mE{{\bm{E}}}

\def\mH{{\bm{H}}}

\def\mM{{\bm{M}}}

\def\mX{{\bm{X}}}

\DeclareMathAlphabet{\mathsfit}{\encodingdefault}{\sfdefault}{m}{sl}
\SetMathAlphabet{\mathsfit}{bold}{\encodingdefault}{\sfdefault}{bx}{n}

\def\gL{{\mathcal{L}}}

\newcommand{\R}{\mathbb{R}}

\makeatletter
\let\save@mathaccent\mathaccent
\newcommand*\if@single[3]{%
  \setbox0\hbox{${\mathaccent"0362{#1}}^H$}%
  \setbox2\hbox{${\mathaccent"0362{\kern0pt#1}}^H$}%
  \ifdim\ht0=\ht2 #3\else #2\fi
  }
\newcommand*\rel@kern[1]{\kern#1\dimexpr\macc@kerna}
\newcommand*\widebar[1]{\@ifnextchar^{{\wide@bar{#1}{0}}}{\wide@bar{#1}{1}}}
\newcommand*\wide@bar[2]{\if@single{#1}{\wide@bar@{#1}{#2}{1}}{\wide@bar@{#1}{#2}{2}}}
\newcommand*\wide@bar@[3]{%
  \begingroup
  \def\mathaccent##1##2{%
    \let\mathaccent\save@mathaccent
    \if#32 \let\macc@nucleus\first@char \fi
    \setbox\z@\hbox{$\macc@style{\macc@nucleus}_{}$}%
    \setbox\tw@\hbox{$\macc@style{\macc@nucleus}{}_{}$}%
    \dimen@\wd\tw@
    \advance\dimen@-\wd\z@
    \divide\dimen@ 3
    \@tempdima\wd\tw@
    \advance\@tempdima-\scriptspace
    \divide\@tempdima 10
    \advance\dimen@-\@tempdima
    \ifdim\dimen@>\z@ \dimen@0pt\fi
    \rel@kern{0.6}\kern-\dimen@
    \if#31
      \overline{\rel@kern{-0.6}\kern\dimen@\macc@nucleus\rel@kern{0.4}\kern\dimen@}%
      \advance\dimen@0.4\dimexpr\macc@kerna
      \let\final@kern#2%
      \ifdim\dimen@<\z@ \let\final@kern1\fi
      \if\final@kern1 \kern-\dimen@\fi
    \else
      \overline{\rel@kern{-0.6}\kern\dimen@#1}%
    \fi
  }%
  \macc@depth\@ne
  \let\math@bgroup\@empty \let\math@egroup\macc@set@skewchar
  \mathsurround\z@ \frozen@everymath{\mathgroup\macc@group\relax}%
  \macc@set@skewchar\relax
  \let\mathaccentV\macc@nested@a
  \if#31
    \macc@nested@a\relax111{#1}%
  \else
    \def\gobble@till@marker##1\endmarker{}%
    \futurelet\first@char\gobble@till@marker#1\endmarker
    \ifcat\noexpand\first@char A\else
      \def\first@char{}%
    \fi
    \macc@nested@a\relax111{\first@char}%
  \fi
  \endgroup
}
\makeatother

\usepackage{pifont}
\newcommand{\xmark}{\ding{55}} 

\usepackage{tikz}
\usepackage{pgfplots}
\pgfplotsset{compat=1.17}
\usepackage{silence}
\usepackage[nameinlink]{cleveref}
\crefname{equation}{Eq.}{Eqs.}
\crefname{figure}{Fig.}{Figs.}
\crefname{section}{Sec.}{Sec.}
\crefname{appendix}{App.}{App.}
\crefname{table}{Tab.}{Tabs.}
\crefname{algorithm}{Algo}{Algo}
\crefname{thm}{Thm}{Thm}
\Crefname{thm}{Thm}{Thm}
\crefname{prop}{Prop}{Prop}

\definecolor{darkred}{rgb}{0.7, 0.0, 0.0}

\newcommand{\audiotoface}{Audio2Face-3D\xspace}

\newcommand{\crefnames}[3]{%
  \@for\next:=#1\do{%
    \expandafter\crefname\expandafter{\next}{#2}{#3}%
  }%
}
\newcommand{\CenterImage}[1]{\raisebox{-.5\height}{#1}} 

\title{\audiotoface: Audio-driven Realistic Facial Animation For Digital Avatars}

\author{NVIDIA\footnote{A detailed list of contributors and acknowledgments can be found in~\cref{sec::contributors} of this paper.}}

\begin{abstract}
Audio-driven facial animation presents an effective solution for animating digital avatars. In this paper, we detail the technical aspects of NVIDIA \audiotoface, including data acquisition, network architecture, retargeting methodology, evaluation metrics, and use cases. \audiotoface system enables real-time interaction between human users and interactive avatars, facilitating facial animation authoring for game characters. To assist digital avatar creators and game developers in generating realistic facial animations, we have open-sourced \audiotoface \href{https://huggingface.co/collections/nvidia/audio2face-3d-6865d22d6daec4ac85887b17}{networks}, \href{https://github.com/NVIDIA/Audio2Face-3D-SDK}{SDK}, \href{https://github.com/NVIDIA/Audio2Face-3D-training-framework}{training framework}, and \href{https://huggingface.co/datasets/nvidia/Audio2Face-3D-Dataset-v1.0.0-claire}{example dataset}. 
\end{abstract}

\begin{document}

\maketitle

\abscontent

\section{Introduction}
\label{sec::intro}

The creation of realistic and interactive digital avatars has become increasingly important in fields such as customer service, gaming, virtual reality, and digital entertainment. High-quality facial animation, essential for achieving lifelike avatars, traditionally involves extensive manual animation and the use of video-based motion capture systems. While video-based motion capture approaches achieve high-quality animation efficiently, these systems require casting skilled actors, capturing their performances with cameras, and handling large volumes of data. This process is not only time-consuming but also costly when applied at scale.

To streamline this process, we present \audiotoface, an advanced audio-driven facial animation system that generates highly realistic facial expressions and synchronized lip movements for digital characters. Audio-driven facial animation enhances the efficiency and accessibility of the animation pipeline by using audio input to drive facial animations, eliminating the need for video-based motion capture.

\audiotoface leverages state-of-the-art deep learning techniques to transform audio input into highly detailed facial animations. By utilizing a high-quality 4D capture dataset and sophisticated network architectures, our system can produce realistic facial animations of skin, tongue, jaw, and eyeballs. The system supports real-time interaction, making it suitable for both live applications and offline facial animation authoring. 

A notable feature of \audiotoface is its capability to adapt to multiple identities and emotional states, offering a comprehensive solution for diverse animation needs. Additionally, we have made our networks, SDK, and training framework open-source, promoting accessibility to advanced digital human technologies and empowering a broader community of developers and creators.

This paper provides an in-depth look at the technical aspects of \audiotoface, including data acquisition, network architectures, retargeting methodologies, evaluation metrics, and practical use cases. We also explore several experimental concepts such as head motion generation, text-conditioned emotion control, and audio-driven rig parameter generation. Our contributions aim to facilitate the creation of realistic digital avatars and democratize the AI-driven facial animation technology.

\section{System Overview}
\label{sec::overview}

\begin{figure}[ht]
    \centering
    \includegraphics[width=\textwidth]{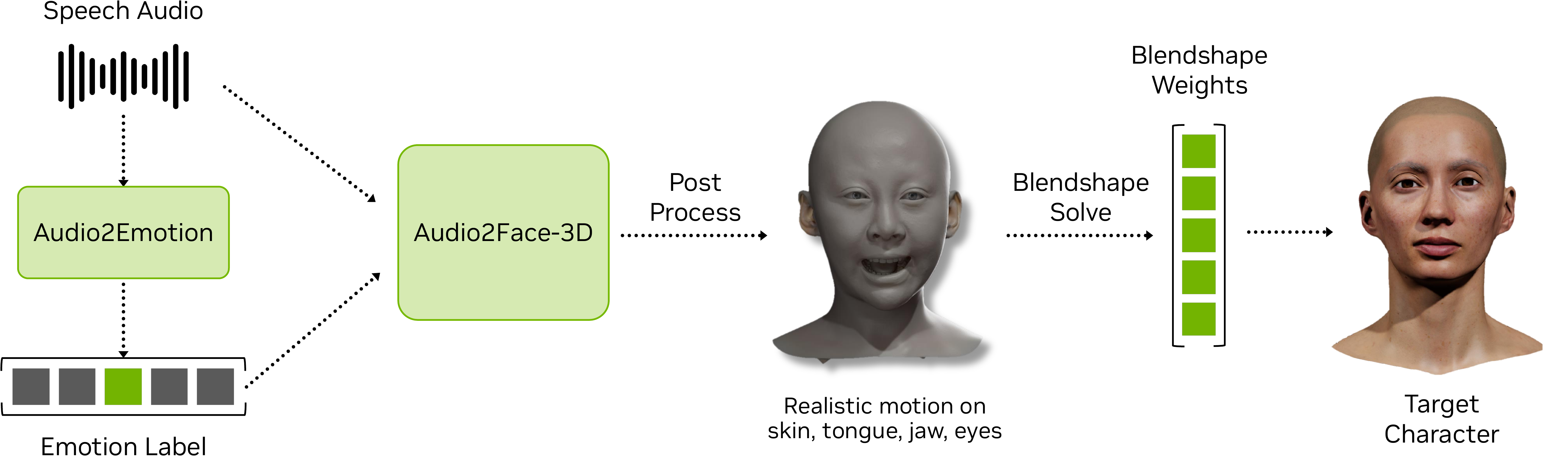}
    \caption{The \audiotoface system receives audio and emotion labels as input and outputs high-fidelity animations for facial skin, tongue, jaw, and eyeballs. To retarget the output facial motion onto a target character, we provide a blendshape solving process. The Audio2Emotion network can optionally provide emotion label over time by detecting emotion from the speech audio.}
    \label{fig:system_overview_a2f}
\end{figure}

In the open-source software, we release two types of \audiotoface networks: 
\begin{itemize}
    \item \emph{\audiotoface-v2.3} networks for individual actors (Mark, Claire, James), referred to as regression-based networks.
    \item \emph{\audiotoface-v3.0} network that supports multiple identities with a single network, based on a diffusion architecture, referred to as a diffusion-based network.
\end{itemize}

Both networks are trained on the same data (\cref{sec::data}) but employ different architectures. The regression-based network (\cref{subsec:regression}) requires only a small chunk (0.52 seconds) of input audio to estimate a facial pose, and its output does not depend on previous or future facial poses. At each inference, the regression-based network outputs a 1-frame pose in a compressed data format. In contrast, the diffusion-based network (\cref{subsec:diffusion}) processes audio in a streaming manner with minimal latency. At each inference, the diffusion-based network uses a 1-second audio chunk and outputs a 30-frame animation in a raw vertex data format. While the regression-based network is lighter on memory and scales well for a large number of concurrent processes, the diffusion-based network generally produces higher quality and more expressive facial animations.

Both networks receive common inputs of audio and emotion labels and output motion deltas for facial skin, tongue, jaw, and eyeballs. Once we obtain the inference output, we post-process it to adjust the scale and offset of each motion component (\cref{sec::postprocess}). After post-processing, we use blendshape decomposition (\cref{sec::bs_solve}) to generate ARKit blendshape~\citep{arkit2025} weight values, which can be used by various face models, such as Epic game's Metahuman or custom blendshape face rigs.

For emotion control, the Audio2Emotion (\cref{sec::audio2emotion}) network can optionally detect the emotion in the voice input and feed the emotion vector to the \audiotoface network. \cref{fig:system_overview_a2f} illustrates the overall workflow of the \audiotoface system.

\section{Data Preparation}
\label{sec::data}

\subsection{Data Capture and Processing}
\label{subsec:capture}
In this section, we describe the facial capture process and how the captured data is processed into training parameters for facial skin, tongue, jaw, and eyeballs. We captured synchronized speech audio and facial animations of professional actors using multiple machine vision cameras provided by a commercial facial capture service \citep{di4d}. The actors were instructed to perform a set of sentences that cover most of the English or Mandarin pronunciations in various emotional states. Each actor performed approximately 50 to 70 target sentences, with lengths ranging from 3 to 15 seconds. The captured motions include 11 different emotional states: neutral, amazement, anger, cheekiness, disgust, fear, grief, joy, out-of-breath, pain, and sadness. We instructed the actors to maintain consistent emotional intensity throughout each sequence. In instances where actors exhibited abrupt ease-in and ease-out of emotions, particularly at the beginning and end of sequences, we clipped those portions and only utilized the consistent segments as training data.

\begin{figure}[!t]
    \centering    
    
    \includegraphics[width=0.7\textwidth]{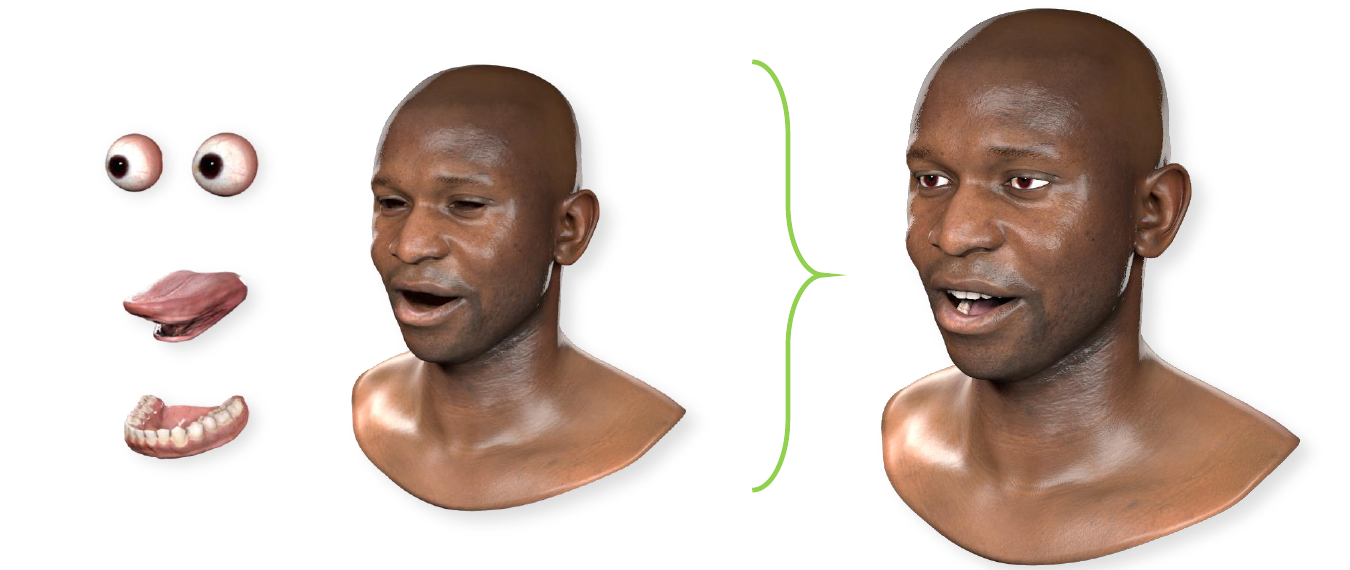} 
    
    
    \caption{Facial components that are used to train \audiotoface animation. The appearance and the motion of each component (skin, tongue, jaw, and eyeballs) is captured from actors and processed to training parameters.}
    \label{fig:data_full_elements}
\end{figure}

Given the basic mesh tracking from the commercial capture service, we convert the tracking data into training parameters (comprising skin, tongue, jaw, and eyeballs in \Cref{fig:data_full_elements}). The resulting training parameters are used to train \audiotoface networks and to generate the realistic facial animation in the inference stage. The following subsections detail the data preparation process for each component. 


\textbf{Skin.} The facial skin capture data is per-frame geometry that tracks the facial deformation using a high-resolution template mesh. Each vertex of the mesh consistently tracks the same skin point of the actor throughout the sequences. To create a compact representation of the data, rather than using the raw high-dimensional vertex representation, we apply Principal Component Analysis (PCA) to the skin data. Specifically, we convert the per-vertex skin deformation of a raw tracked mesh into a lower-dimensional vector representing 3D displacements from the neutral face mesh. This conversion facilitates data uploading and sharing, streamlines data loading during network training, and reduces the number of parameters in the network. We use 140-dimension vector to represent the facial skin deformation, with the exception of v2.3-Mark network (272-dimension). 

\textbf{Tongue.} Since the tongue is not visible in a vision-based capture system, accurate tracking data for the tongue is not available from the commercial capture service. To obtain training data for the tongue, a professional animator manually animated the tongue based on the tracked jaw motion, ensuring correct phonetic characteristics and photo-consistency by referencing the audio and video. 

Once we had a complete set of tongue animations for one capture subject, we trained a temporary \audiotoface network for that actor. This pre-trained network could then generate new tongue motions given new audio input from other actors. After generating the new tongue motion, we apply a simple affine transform to adapt the motion to the target actor's tongue shape. We found that this synthetic data generation method scales well for new capture subjects, and the resulting tongue motion exhibits realistic tongue deformation. After collecting the tongue animations from all training sequences of each actor, we applied PCA to the tongue data to create a compact 10-dimensional vector representation.

\textbf{Jaw.} The jaw (and lower teeth) geometry are tracked by the commercial capture service. Since the jaw is a rigid objects, its per-frame motion can be represented as a rigid transformation matrix. We select five mesh vertices from the geometry (left-back, right-back, left-front, right-front, and center-front vertices) that can reliably provide the rigid transform and record their per-frame 3D displacement from the neutral position. This per-frame 15-dimensional data, along with the neutral positions of these vertices, is used as a trainable representation of the jaw motion.

\textbf{Eyeballs.} The eyeball rotation is tracked by the commercial service. We further refine this tracked data to re-compute clean yaw and pitch rotations within the eyeball rig space. The reconstructed eyeball motion accurately captures eye movements, including gaze, idle, and saccade motions. The eyeball rotation is represented as a 4-dimensional vector, utilizing two Euler angles for each of the left and right eyeballs relative to their neutral rotation.


\begin{figure}[!t]
    \centering
    \includegraphics[width=\textwidth]{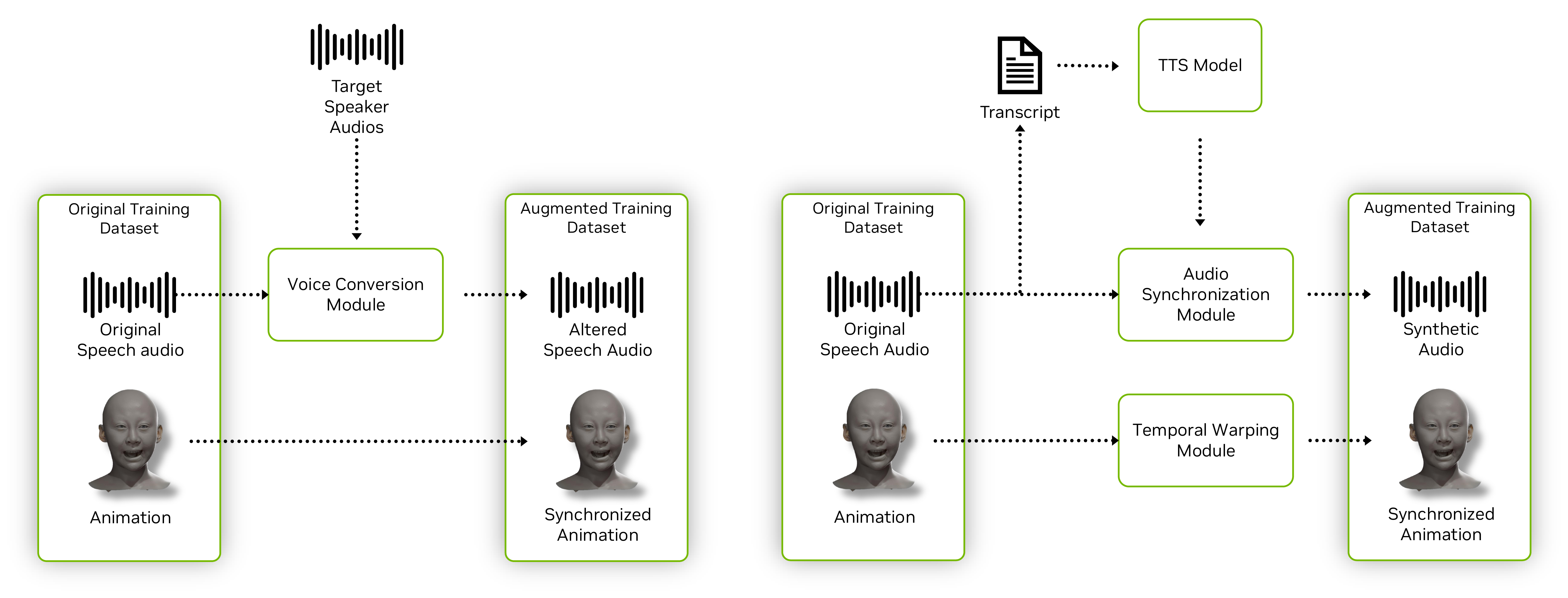}
    \caption{Dataset augmentation strategies in \audiotoface. \emph{left:} Voice cloning based augmentation. \emph{right:} Text-to-speech and temporal animation warping augmentation }
    \label{fig:data_full_augmentations}
\end{figure}

\subsection{Dataset Augmentation}
\label{subsec:data_aug}
Datasets for audio-driven 3D facial animation primarily consist of paired audio and 4D facial animation data. However, collecting these data is labor intensive and costly and often lacks the diversity of speech and expression required for training robust models. Basic augmentations, such as pitch shifting or temporal stretching, fail to produce realistic voice variation or easily break the alignment with facial motion. To address these limitations, we employ three augmentation techniques that generate new synchronized audio-visual training samples from existing data.

\textbf{Voice Conversion.} 
Voice conversion~\citep{azzuni2025voicecloning,dhar2025ganvc} modifies an audio recording to simulate the realistic voice of a different speaker by altering acoustic features such as pitch, timbre, and formants, while preserving timing characteristics such as speech rhythm and duration. Since the temporal structure of the speech remains unchanged, the original facial animation remains synchronized, allowing new voice variants to be generated without modifying the animation.

\textbf{Text-to-Speech (TTS) Synthesis.} 
TTS synthesis creates new audio samples from the transcripts of original recordings. Unlike voice conversion, TTS may introduce additional variations in timing, rhythm, and prosody. To maintain synchronization, we first apply phoneme detection~\citep{zhu2022Charsiu} to both original and TTS audio. Then, a temporal alignment step such as Dynamic Time Warping (DTW)~\citep{sakoe1978dtw} is performed, guided by the detected phoneme boundaries, to adjust the timing of the original animation to match the new speech signal.

\textbf{Silence data.}
Silence is a common input for audio-driven facial animation, particularly at the beginning or end of an audio clip. In many cases, users expect minimal and stable motion when the input is silent. To achieve this, we augment each emotional state with 4 seconds of silence data, using silent audio and a silent expression from the capture data. When the network is trained with these augmented data, it can better achieve the desired expressions for silent audio, such as closed lips or a natural smile.

These augmentation strategies expand the diversity of training data while preserving synchronization. The resulting samples can be incorporated into the original dataset, improving the model’s ability to generalize across voice identities, speaking styles, and prosodic variation.

\section{\audiotoface Network}
\label{sec::models}
Our open-weight \audiotoface networks include two distinct architectures: a regression-based network (\audiotoface-v2.3) and a diffusion-based network (\audiotoface-v3.0), which will be described in \cref{subsec:regression} and \cref{subsec:diffusion}, respectively.



\subsection{\audiotoface-v2.3: Regression-based Network}
\label{subsec:regression}
\begin{figure}[!t]
    \centering
    \includegraphics[width=0.7\textwidth]{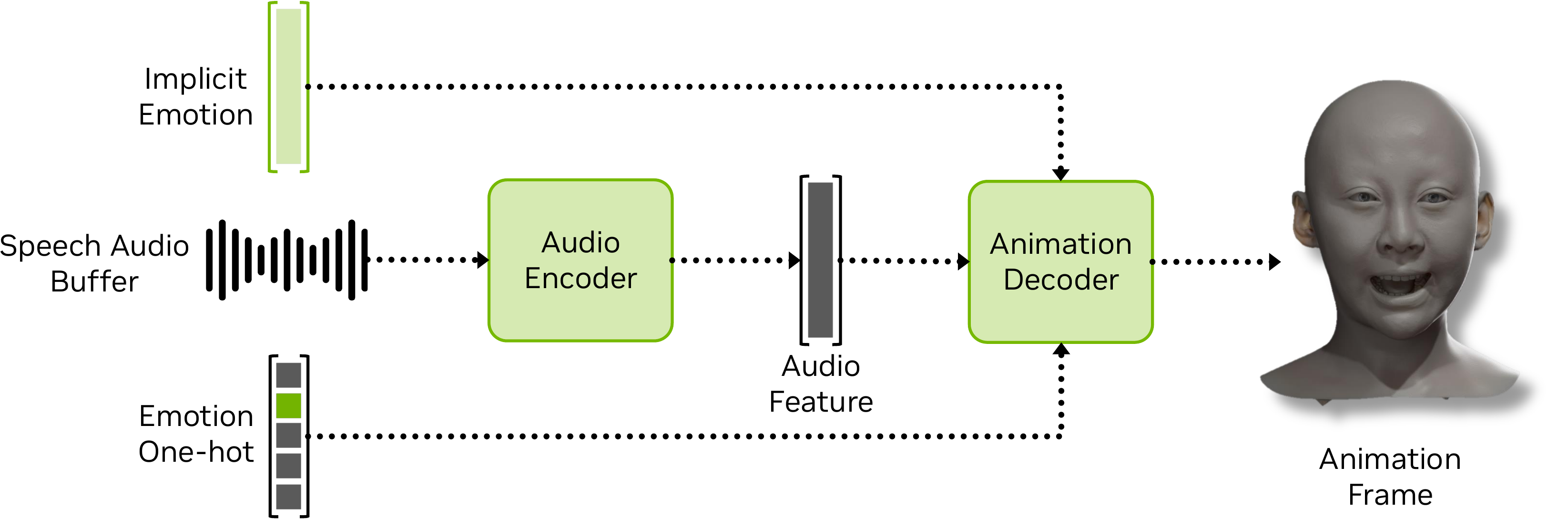}
    \caption{Regression network architecture. The network extracts the audio feature from the audio encoder. The extracted audio feature, the implicit emotion, and one-hot emotion are then fed to the animation decoder to get the animation frame. }
    \label{fig:regression_network}
\end{figure}

The regression network has a functional form of $f_{{\theta}} (\mA, \ve) \rightarrow (\vx_{\text{skin}}, \vx_{\text{tongue}}, \vx_{\text{jaw}}, \vx_{\text{eye}}) := \vx$, where $\mA$ is speech audio, $\ve$ is an emotion vector, $\vx_{\text{skin}}$ is facial skin pose, $\vx_{\text{tongue}}$ is tongue pose, $\vx_{\text{jaw}}$ is jaw pose, and $\vx_{\text{eye}}$ is eye rotation. The network is based on the prior work~\citep{karras2017afa}, but with some important modifications on the network architecture and the loss function to improve the network's performance. The reader can refer to the original paper for more details. 

\noindent\textbf{Network Architecture.} The network architecture consists of several key components: the audio encoding module, the animation decoding module, and the phoneme prediction module. For the audio encoding module, instead of using an autocorrelation-based audio encoder as in~\citep{karras2017afa}, we use a hybrid audio encoder that combines autocorrelation features and Wav2Vec 2.0 features~\citep{baevski2020wav2vec}. Wav2Vec 2.0 is a self-supervised learning model that learns speech representations from unlabeled audio data, and incorporating it into the network improves the network's lip sync quality and multi-lingual capabilities. At the same time, we found that the autocorrelation features are still useful in extracting pitch and volume-related information in the audio, which improves the network's performance in singing and some non-verbal audio. 

The phoneme prediction module is designed to improve lip-sync accuracy by explicitly predicting phonemes from the audio features as an auxiliary task. This module takes the extracted audio features from the audio feature encoding module and applies a lightweight convolutional head to predict phoneme probabilities. This helps the network learn more accurate mouth shapes for bilabial sounds, as the phoneme prediction task enforces understanding the phonemes being spoken in the audio, which is also beneficial for facial animation generation. Ground truth phonemes are generated at 50 fps for each track before training and then later used during training. The phoneme prediction module is trained jointly with the rest of the network, using the cross-entropy loss. It is later dropped when the whole network is trained, as its main purpose is to propagate phoneme-associated information through gradients to the audio encoder.

\begin{figure}[!t]
    \centering
    \includegraphics[width=0.85\textwidth]{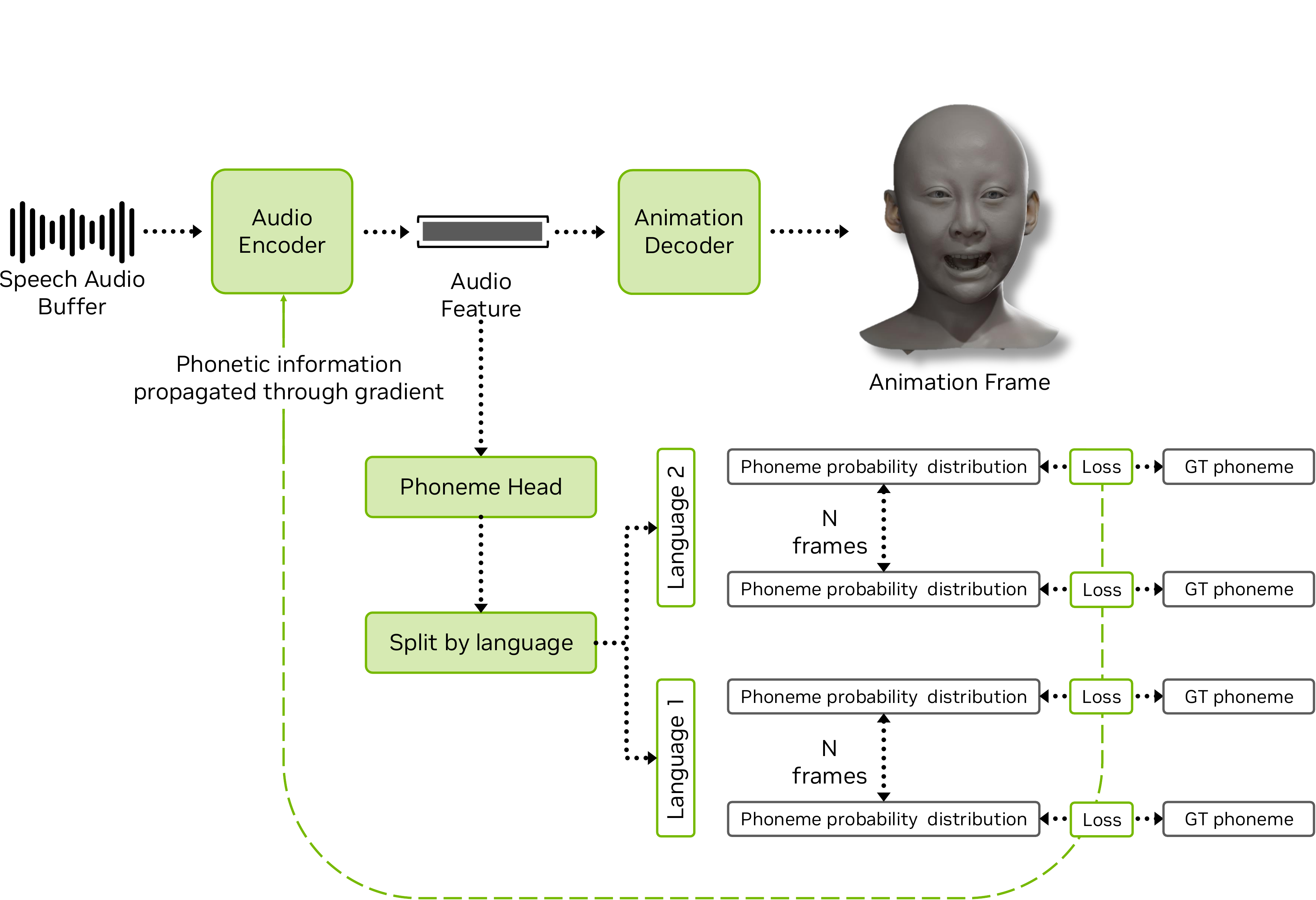}
    \caption{Illustration of the phoneme prediction head. It predicts the phoneme probabilities for the input audio. Separate phoneme heads are used for each language. The cross entropy loss is applied on the phoneme head and is trained jointly with the rest of the model.}
    \label{fig:phoneme_head}
\end{figure}

The animation decoding module consists of a set of convolutional and linear layers. It takes the extracted audio feature from the audio encoding module, emotion vector $\ve$, to generate the full-face pose $\vx$. The emotion vector $\ve = \text{concat}(\ve^{\text{explicit}}, \ve^{\text{implicit}})$, where $\ve^{\text{explicit}}$ is the one-hot encoded explicit emotion vector (joy, anger, etc.) and $\ve^{\text{implicit}}$ is the implicit emotion vector learned during training to model the frame-to-frame temporal variation that are not explained by the audio and explicit emotion~\citep{karras2017afa}.

\noindent\textbf{Loss Function.}
The regression-based network employs a specialized loss function that combines several objectives to ensure high-quality facial animation generation. The total loss $\mathcal{L}_{\text{total}}$ is formulated as:

\begin{align}
\mathcal{L}_{\text{total}} &= \alpha_{\text{mse}} \mathcal{L}_{\text{mse}} + \alpha_{\text{motion}} \mathcal{L}_{\text{motion}} + \alpha_{\text{phoneme}} \mathcal{L}_{\text{phoneme}} \nonumber\\
&\quad + \alpha_{\text{phoneme\_motion}} \mathcal{L}_{\text{phoneme\_motion}} + \alpha_{\text{vol\_stab}} \mathcal{L}_{\text{vol\_stab}} + \alpha_{\text{emo}} \mathcal{L}_{\text{emo}} \nonumber\\
&\quad + \alpha_{\text{lip\_dist}} \mathcal{L}_{\text{lip\_dist}} + \alpha_{\text{lip\_size}} \mathcal{L}_{\text{lip\_size}}.
\end{align}

The first component $\mathcal{L}_{\text{mse}} =\|\hat{\vx} - \vx\|_2^2$ is a mean square error term between predicted full-face coefficients $\hat{\vx}$ and ground truth full-face coefficients $\vx$. The motion loss term $\mathcal{L}_{\text{motion}} = \|(\hat{\vx}_{n+1} - \hat{\vx}_n) - (\vx_{n+1} - \vx_n)\|_2^2$ encourages temporal stability by minimizing velocity error. For improved bilabial sound lip shapes, we include a phoneme prediction loss $\mathcal{L}_{\text{phoneme}} = -\sum_{n,c} p_{n,c} \log(\hat{p}_{n,c})$ using cross-entropy between ground truth and predicted phoneme probabilities $p_{n,c}$ and $\hat{p}_{n,c}$. Additionally, we also apply motion loss term $\mathcal{L}_{\text{phoneme\_motion}} = \|(\hat{p}_{n+1} - \hat{p}_n) - (p_{n+1} - p_n)\|_2^2$ to the predicted phoneme probabilities to ensure smooth phoneme transitions over time. 

The volume-based stability regularization $\mathcal{L}_{\text{vol\_stab}}$ reduces temporal jitter during quiet audio segments:
\begin{equation}
\mathcal{L}_{\text{vol\_stab}} = w_{\text{vol}} \cdot \|\hat{\vx}_{n+1} - \hat{\vx}_n\|_2^2,
\end{equation}
where $w_{\text{vol}} = \exp(-\beta \cdot \text{volume}(\mA))$ provides higher regularization weight during low-volume segments. 

Emotion regularization $\mathcal{L}_{\text{emo}}$ applies temporal smoothness constraints on the implicit emotion vectors $\ve^{\text{implicit}}$:
\begin{equation}
\mathcal{L}_{\text{emo}} = \frac{\|\ve^{\text{implicit}}_{n+1} - \ve^{\text{implicit}}_n\|_2^2}{\|\ve^{\text{implicit}}_n\|_2^2},
\end{equation}
where $\ve^{\text{implicit}}_n$ represents the learnable implicit emotion vector associated with frame $n$.

For realistic lip movements, we employ two specialized lip losses. The lip distance loss $\mathcal{L}_{\text{lip\_dist}}$ constrains the distance between corresponding upper and lower lip key points on the face mesh:
\begin{equation}
\mathcal{L}_{\text{lip\_dist}} = \sum_{i=1}^{5} w_i \cdot \|(\hat{d}^{\text{upper}}_i - \hat{d}^{\text{lower}}_i) - (d^{\text{upper}}_i - d^{\text{lower}}_i)\|_2^2,
\end{equation}
where $\hat{d}^{\text{upper}}_i$ and $\hat{d}^{\text{lower}}_i$ are $i$-th upper and lower lip key points on the predicted mesh, $d^{\text{upper}}_i$ and $d^{\text{lower}}_i$ are key points on the ground truth mesh, and $w_i = \exp(-\gamma \cdot \max(d^{\text{upper}}_i - d^{\text{lower}}_i, 0))$ provides adaptive weighting that emphasizes smaller lip openings.

The lip size loss $\mathcal{L}_{\text{lip\_size}}$ encourages the volume of the predicted upper and lower lips to be close to the ground truth (prevent unnatural lip thinning):
\begin{equation}    
\mathcal{L}_{\text{lip\_size}} = \sum_{i=1}^{5} \|\hat{h}^{\text{upper}}_i - h^{\text{upper}}_i\|_2^2 + \|\hat{h}^{\text{lower}}_i - h^{\text{lower}}_i\|_2^2,
\end{equation}
where $\hat{h}^{\text{upper}}_i$ and $\hat{h}^{\text{lower}}_i$ represent the predicted thickness of the $i$-th upper and lower lip segments, and $h^{\text{upper}}_i$ and $h^{\text{lower}}_i$ are the corresponding ground truth thicknesses. 

\noindent\textbf{Implementation Detail.}
We train the model using Adam optimizer with an initial learning rate of $2 \times 10^{-4}$, scheduled using step decay with $\gamma = 0.994$ per epoch. The model is trained on an NVIDIA RTX A6000 for 50 epochs (28 minutes) for Claire and Mark, and 70 epochs (40 minutes) for James, with a batch size of 32. The loss weights are identity-specific: for James, $\alpha_{\text{mse}} = 1.0$, $\alpha_{\text{motion}} = 10.0$, $\alpha_{\text{phoneme}} = 0.1$, $\alpha_{\text{phoneme\_motion}} = 0.1$, $\alpha_{\text{vol\_stab}} = 100.0$, $\alpha_{\text{emo}} = 1.0$, $\alpha_{\text{lip\_dist}} = 50.0$, and $\alpha_{\text{lip\_size}} = 0.01$. For Claire, $\alpha_{\text{mse}} = 1.5$, $\alpha_{\text{motion}} = 10.0$, $\alpha_{\text{phoneme}} = 0.2$, $\alpha_{\text{phoneme\_motion}} = 0.1$, $\alpha_{\text{vol\_stab}} = 100.0$, $\alpha_{\text{emo}} = 1.0$, and $\alpha_{\text{lip\_dist}} = 100.0$. For Mark, $\alpha_{\text{mse}} = 1.0$, $\alpha_{\text{motion}} = 10.0$, $\alpha_{\text{phoneme}} = 1.0$, $\alpha_{\text{phoneme\_motion}} = 0.0$, $\alpha_{\text{vol\_stab}} = 100.0$, $\alpha_{\text{emo}} = 1.0$, and $\alpha_{\text{lip\_dist}} = 500.0$. During training, similar to prior work~\citep{karras2017afa}, each sample in the batch is a temporal pair consisting of two adjacent frames. This facilitates the loss calculation that involves temporal smoothness, i.e., $\mathcal{L}_{\text{motion}}$, $\mathcal{L}_{\text{phoneme\_motion}}$, $\mathcal{L}_{\text{vol\_stab}}$. We use textless phoneme aligner~\citep{zhu2022Charsiu} to generate ground truth phoneme probability labels using training audios.


 We implement eye-closure filtering based on the ground truth eye-opening size to remove training samples with closed eyes, which prevents the model from generating unstable eyelid movements. We use the audio feature after the 4th transformer layer of Wav2Vec2 with frozen weights. Input audio $\mA$ is at 16kHz sampling rate and has a fixed size of 8320 samples. For each input audio buffer, the model outputs 169-dimensional full-face coefficients $\hat{\vx}$, including 140 skin PCA dimensions, 10 additional tongue PCA coefficients, 15 jaw dimensions, and 4 eyeball rotation dimensions. To obtain the final full-face pose $\vx$, we apply the inverse PCA transformation to the PCA coefficients. During inference, we use a sliding window approach to generate the animation frame by frame, with an audio sample stride of $s =16000/F$, where $F$ is the desired frame rate.


\subsection{\audiotoface-v3.0: Diffusion-based Network}
\label{subsec:diffusion}
Diffusion probabilistic network~\citep{ho2020denoising} learns to generate data from a target distribution by progressively denoising a normally distributed variable, demonstrating promising performance in both 2D~\citep{rombach2022stable, saharia2022imagen, ramesh2022dalle2} and 3D generation~\citep{stan2023facediffuser}. 
We employ a diffusion mechanism to generate high-quality and diverse facial animations.

\begin{figure}[h]
    \centering
    \includegraphics[width=\textwidth]{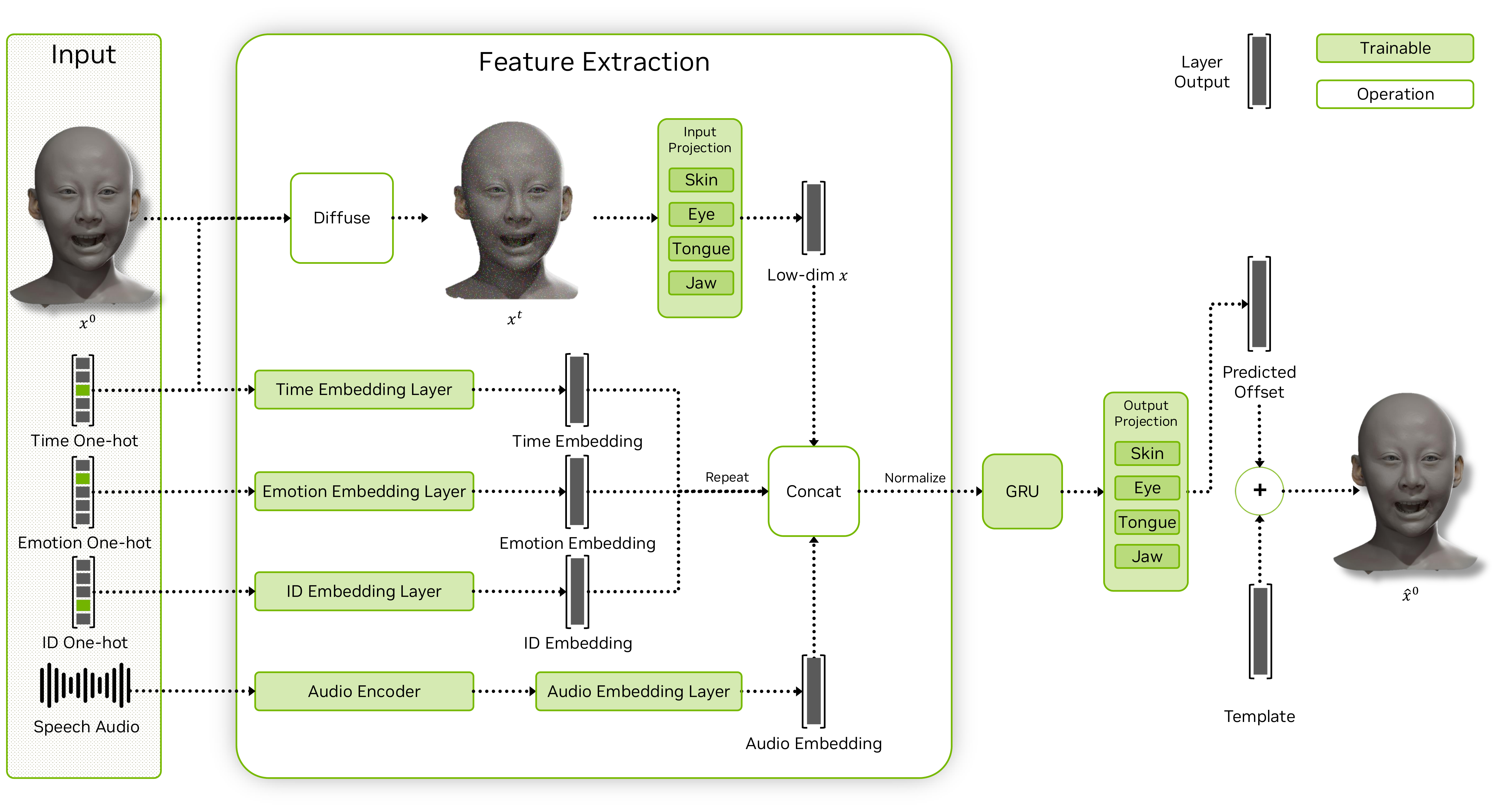}
    \caption{Diffusion-based network architecture. It denoises the input animation using a GRU based structure, conditioned on the diffusion timestep, emotion, identity and audio. The network predicts the offset, which is added to the neutral face template to get the denoised animation.}
    \label{fig:diffusion_network}
\end{figure}

\noindent\textbf{Network Architecture.} The diffusion-based network architecture is illustrated in \cref{fig:diffusion_network}. The diffusion-based network not only supports emotion conditioning but also identity conditioning using a one-hot identity vector.
This enables the generation of facial animations for multiple identities with a single network.
Specifically, the diffusion network predicts $N$ frames of denoised facial animation as follows:  
\begin{equation}
    \hat{\mX}_0 = f_\theta(\mA,\mX_t, \vt, \mE, \vi),
\end{equation}
where $f_\theta$ is the parameterized model, $\mA$ is the input audio, $\mX_t \in \R^{N \times D_x}$ is the noisy facial animation, $\vt \in \R$ is the timestep, $\mE \in \R^{N \times D_e}$ is the per-frame emotion vector, $\vi$ is the identity vector, and $\hat{\mX}_0 \in \R^{N \times D_x}$ is the predicted facial animation.
We first encode audio $\mA$ using HuBERT~\citep{hsu2021hubert}, a large-scale pretrained speech representation model, then each input (HuBERT audio feature, noisy animation, timestep, emotion, and identity) is passed through distinct projection layers, and the resulting outputs are concatenated before being fed into GRU layers, which are well-suited for modeling temporal sequences.
Finally, the decoding module decodes the GRU output to generate the denoised full-face animation $\hat{\mX}_0$.

\noindent\textbf{Loss Function.}
We employ a simple objective function $\mathcal{L}_{\text{simple}}$ to train the diffusion-based network as in the prior work~\citep{ho2020denoising}. 
Unlike the standard approach that predicts noise $\epsilon$, our network directly predicts the denoised facial animation $\hat{\mX}_0$ from the noisy input $\mX_t$. We apply a mean square error loss to the denoised output $\hat{\mX}_0$ to compute $\mathcal{L}_{\text{simple}}$, following the previous work~\citep{ramesh2022dalle2}, as:
\begin{equation}
    \mathcal{L}_{\text{simple}} = \mathbb{E}_{\mX_0\sim q(\mX_0|c),t\sim[1,T]}\left[\lVert\mX_0 - \hat{\mX}_0\rVert_F^2\right],
\end{equation}
where $c = \{\mA, \mE, \vi\}$ represents the conditioning information, including audio, emotion, and identity.
Also, we use lip distance loss $\mathcal{L}_{\text{lip\_dist}}$ as in the regression-based network for better bilabial sound lip shapes.
Additionally, we stabilize upper face motion and prevent temporal jittering with an expression smoothness regularization loss $\mathcal{L}_{\text{upper\_reg}}$.
This loss encourages temporal coherence by minimizing inter-frame variations in the upper facial region:
\begin{equation}
    \mathcal{L}_{\text{upper\_reg}} = \|\hat{\mX}_{0,n+1}^{\text{upper\_face}} - \hat{\mX}_{0,n}^{\text{upper\_face}}\|_F^2,
\end{equation}
where $\hat{\mX}_{0,n}^{\text{upper\_face}}$ represents the denoised upper face animation segment starting at frame $n$.
The total loss function is given by:
\begin{equation}
    \mathcal{L}_{\text{total}} = \mathcal{L}_{\text{simple}} + \alpha_{\text{upper\_reg}}\mathcal{L}_{\text{upper\_reg}} + \alpha_{\text{lip\_dist}}\mathcal{L}_{\text{lip\_dist}}.
\end{equation}
 
\noindent\textbf{Implementation Detail.}
During training, we randomly sample a timestep $\vt \in \{1, ..., T\}$ and directly sample the noisy facial animation $\mX_t$ from $\mX_0$ using a predefined noise schedule as follows:
\begin{equation}
    \mX_t = \sqrt{\bar{\alpha}_t}\mX_0 + \sqrt{1-\bar{\alpha}_t}\epsilon,
\end{equation}
where $\epsilon \sim \mathcal{N}(0, \mathbf{I})$ is sampled Gaussian noise, and $\bar{\alpha}_t = \prod_{s=1}^{t}\alpha_s$ with $\alpha_t = 1 - \beta_t$ and $0<\beta_1 < ... < \beta_T < 1$.
We use Adam optimizer with an initial learning rate of $5 \times 10^{-5}$, scheduled using cosine annealing with a minimum learning rate of $10^{-5}$ and a warm-up period of 10 epochs. The network is trained on an NVIDIA RTX A6000 for 400 epochs (5 hours) with a batch size of 1, using gradient accumulation over 4 steps. The loss weights are $\alpha_{\text{upper\_reg}} = 25$ and $\alpha_{\text{lip\_dist}} = 1 \times 10^{-5}$. We set the number of diffusion steps $T=1000$ and use a cosine noise schedule. During training, we apply data augmentation, including random audio pitch shifting and variable-length audio segment extraction (30-600 frames). Training using variable-length audio segments was found to be useful for improving the performance of the network when using streaming inference, which will be described next.

\begin{figure}[h]
    \centering
    \includegraphics[width=0.65\textwidth]{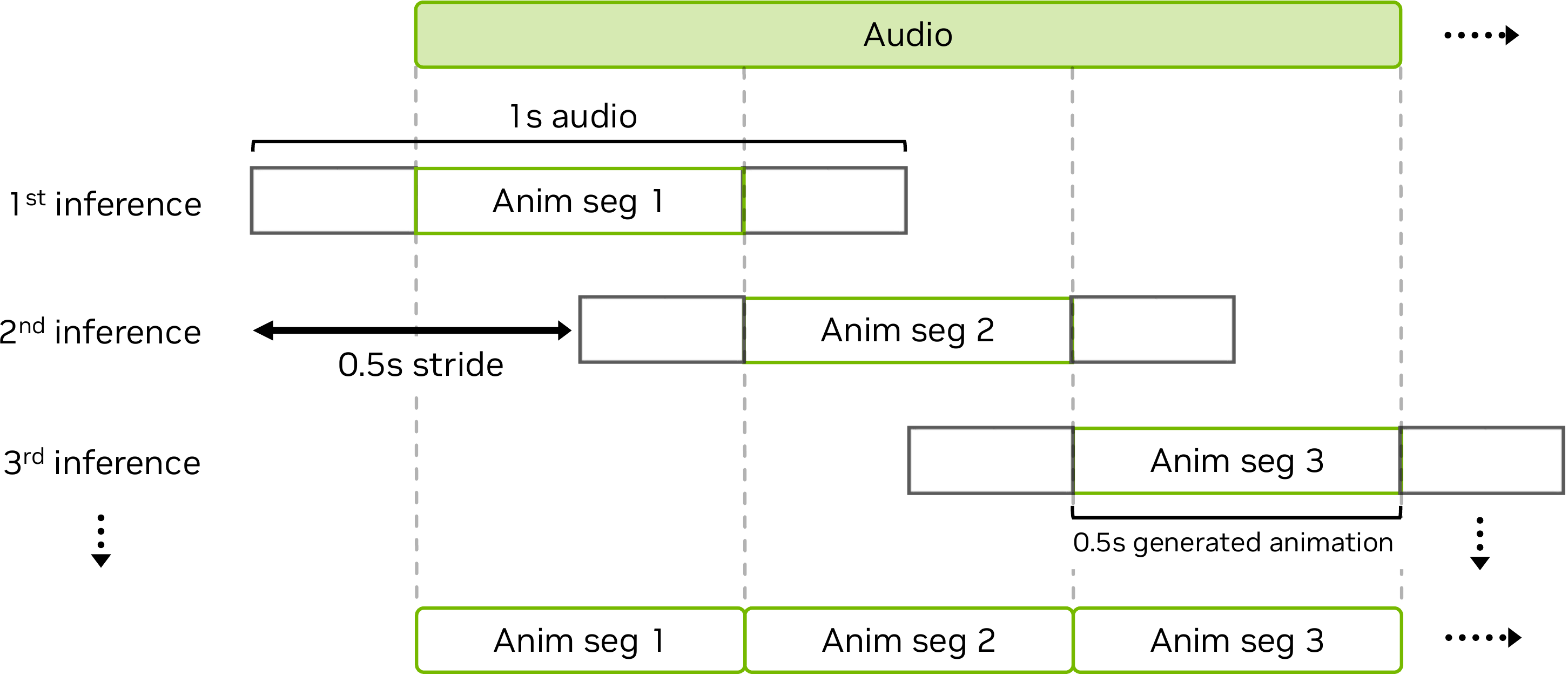}
    \caption{Illustration of streaming inference. The network extracts the audio feature from the 1s audio window and generates the central 0.5s animation segment. This process repeats with a stride of 0.5s and generates the animation in streaming mode.}
    \label{fig:streaming_inference}
\end{figure}
\noindent\textbf{Inference.} The proposed diffusion-based network supports both offline and real-time streaming inference modes. For offline inference, we denoise the entire audio sequence at once using the reverse diffusion process. For real-time applications, we implement a streaming inference approach as illustrated in \cref{fig:streaming_inference}, where audio segments are processed in a sliding window manner. During each inference step, the network uses audio features extracted from a 1-second audio segment along with the GRU hidden states from the previous step. The network's denoising process then generates the central 0.5-second facial animation segment and updates the GRU hidden states. This process repeats with an audio segment stride of 0.5~seconds, enabling continuous streaming generation of the entire facial animation. We find that using 2 diffusion steps is sufficient to generate high-quality facial animations, significantly reducing the inference time.


\newpage

\section{Post-Processing}
\label{sec::postprocess}

\begin{figure}[!t]

    \centering
    
    \includegraphics[height=0.32\textwidth]{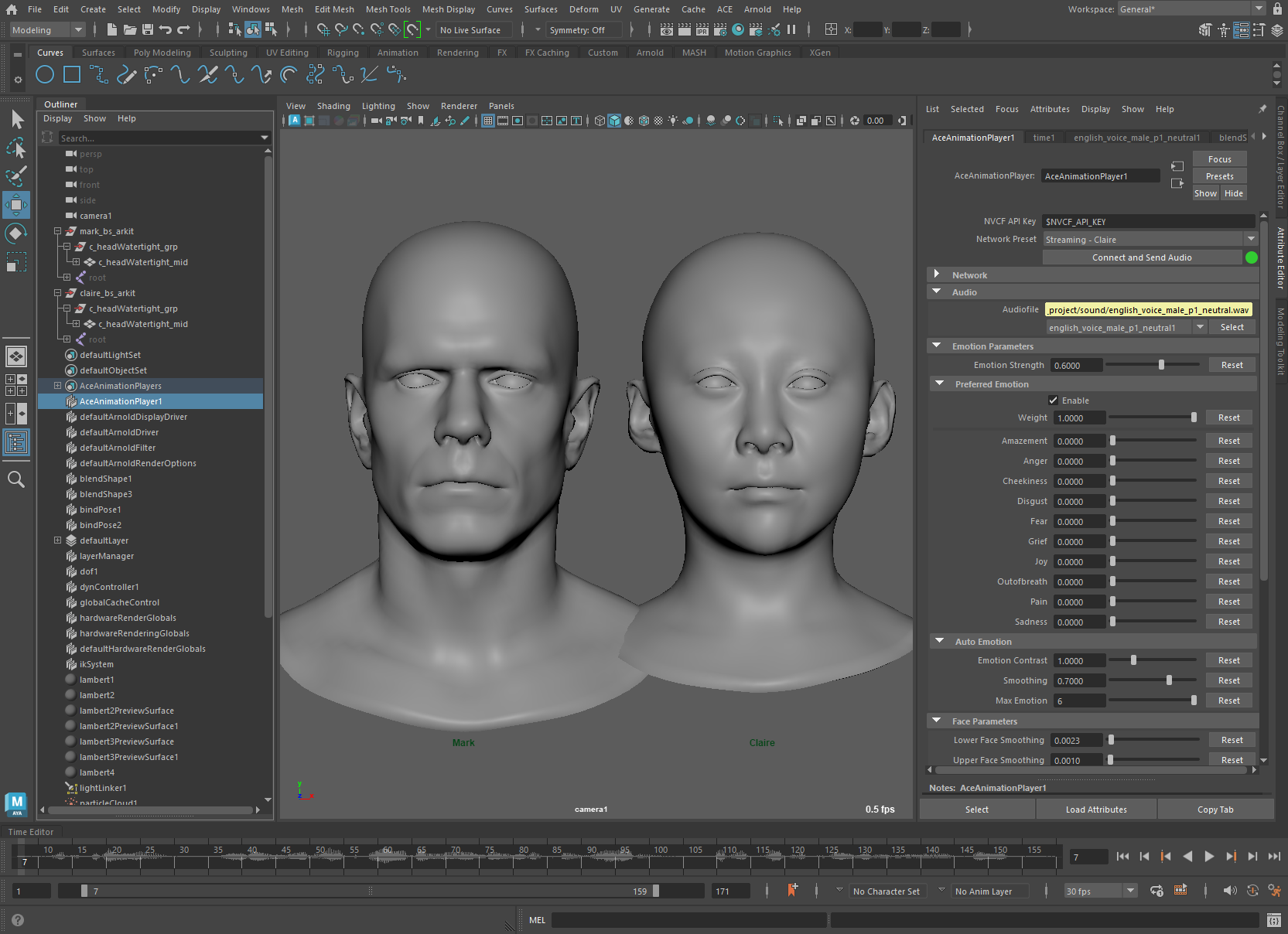}
    \hspace{0.01\textwidth}
    \includegraphics[height=0.32\textwidth]{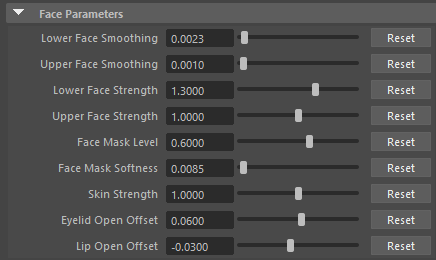}

    \caption{
        A2F postprocessing in Maya-ACE. \emph{Left:} The user interface of the ACE plugin for Maya. \emph{Right:} The post-processing parameters that allow the user to refine the inference output.
    }
    \label{fig:postprocess}

\end{figure}

\audiotoface includes a rich post‑processing suite that enables users to refine facial animation output after neural inference. These adjustments allow for per-region expression tuning, motion smoothing, and anatomical corrections, providing a degree of control without re-training or modifying the network (\cref{fig:postprocess}).

\textbf{Skin Controls.} These parameters control the overall dynamics of facial motion:
\begin{itemize}[itemsep=0.3\baselineskip]
    \item \textbf{Skin Strength:} Controls the overall range of motion of the entire face region. 
    \item \textbf{Upper / Lower Face Strength:} Control the range of the motion of the upper and lower face regions.
    \item \textbf{Upper / Lower Face Smoothing:} Reduces the temporal jitter of the upper and lower regions of the face. Previous frames are sequentially blended into the current frame with decaying weights.
    \item \textbf{Face Mask Level / Softness:} Controls the spatial transition between upper and lower face filters.
\end{itemize}   

\textbf{Jaw and Tongue Controls.} Parameters to control the behavior of the jaw and tongue:
\begin{itemize}[itemsep=0.3\baselineskip]   
    \item \textbf{Lip Open Offset:} Offsets the resting pose of the lips.
    \item \textbf{Jaw Strength, Height, Depth:} Controls the range of motion and positioning of the jaw.
    \item \textbf{Tongue Strength, Height, Depth:} Controls the range of motion and positioning of the tongue.
\end{itemize}    

\textbf{Eye Controls.} Parameters for managing blink and gaze dynamics:
\begin{itemize}[itemsep=0.3\baselineskip]
    \item \textbf{Eyelid Offset:} Offsets the default pose of the eyelid.
    \item \textbf{Blink Strength:} Controls the range of motion of the eye blink.
    \item \textbf{Offset Strength:} Controls the range of motion of the eye offset per emotion.
    \item \textbf{Saccade Strength:} Controls the range of motion for the eye darting behavior.
    \item \textbf{Eye Rotation (X, Y):} Offsets the orientation of the right and left eyes.
\end{itemize}

\newpage
\section{Blendshape Solve}
\label{sec::bs_solve}

\audiotoface generates facial animation by directly driving vertex positions from audio input. While this produces expressive and detailed results, the generated motion is tied to a specific \audiotoface face identity. This limits its direct use in downstream applications, where users typically want to animate their own characters.

Blendshapes, or morph targets, are a standard approach to facial rigging, representing facial expressions as linear combinations of pre-defined deformation targets. To support this, we provide a general blendshape solver that fits a blendshape model to the vertex motion generated by \audiotoface. In addition, we release an ARKit-compatible model to help users obtain the corresponding weights and integrate \audiotoface more easily into their workflows.

\textbf{Delta Blendshape Formulation.}
 We adopt the \emph{delta-blendshape} formulation~\cite{Lewis2014}, where each blendshape encodes an offset (\emph{delta}) from the neutral expression. Let $\vv \in \R^{3V}$ denote a facial mesh, where $V$ is the number of vertices and each vertex has 3 coordinates (X, Y, Z). Given a set of $N+1$ expression meshes $\vv_0, \vv_1, \dots, \vv_N$, with $\vv_0$ denoting the neutral face, a facial expression is defined as: 
 
\begin{equation}
\vv \approx \vv_0 + \sum_{i=1}^{N} w_i \Delta \vv_i = \vv_0 + \mD \vw,
\end{equation}

where $\vw \in [0, 1]^N$ is the blendshape weight vector and $\mD = \begin{bmatrix} \Delta \vv_1 & \Delta \vv_2 & \cdots & \Delta \vv_N \end{bmatrix} \in \mathbb{R}^{3V \times N}$ assembles the \emph{delta-blendshapes} $\Delta \vv_i := \vv_i - \vv_0$ in matrix form. The optimization is performed in delta space, fitting the displacement $\Delta \vv := \vv - \vv_0$ to the weighted blendshape displacements $\mD \vw$.

\textbf{Vertex Sampling.}
To reduce computation and focus on visually salient regions we fit the model on a subset of facial vertices $\vv_{s} \in \R^{3 V_s}$ by applying a sparse selector matrix $\mM \in \R^{3 V_s \times 3V}$ to both the target displacements and blendshape displacements, fitting $\mM\Delta \vv \approx \mM\mD\vw$. We sample only the frontal region, exclude non-expressive areas such as the scalp and shoulders, and increase sampling density in perceptually sensitive areas like the lips, eyelids, and nasolabial folds. For clarity, we omit $\mM$ in subsequent equations.

\subsection{Optimization Objective}

Given a target expression mesh, our goal is to compute the blendshape coefficients that produce a geometry closely approximating the target. Specifically, we aim to estimate blendshape weights $\vw$ that minimize the reconstruction error between the observed and synthesized facial geometry in delta space, subject to box constraints on $\vw$. To ensure stability, sparsity, and temporal coherence, we incorporate regularization terms into the objective. The resulting optimization problem is:

\begin{equation}
\min_{\vw \in [0, 1]^N} 
\underbrace{\left\| \Delta \vv - \mD \vw \right\|_2^2}_{\text{data term}}
+ \underbrace{ \lambda_{L2} \left\| \vw \right\|_2^2}_{\text{ $\mathbf{L}_2$ regularization}}
+ \underbrace{\lambda_{L1} \left\| \vw \right\|_1}_{\text{sparsity ($\mathbf{L}_1$)}}
+ \underbrace{\lambda_{T} \left\| \vw - \vw_{\text{prev}} \right\|_2^2}_{\text{temporal smoothness}}
\end{equation}

Each regularization term is weighted by a tunable coefficient.

\textbf{L2 Regularization (Weight Penalty).} This term discourages large weight magnitudes and helps avoid overfitting. It favors distributed, low-magnitude weights, which are particularly helpful in ill-posed or ambiguous cases.

\textbf{L1 Regularization (Sparsity).} To promote sparse activation patterns where only a few blendshapes are used per frame, we include an $\mathbf{L}_1$ regularizer. This is implemented via a quadratic proxy $\vw^\top \mathbf{L}_1 \vw$, where $\mathbf{L}_1 = \mathbf{1} \cdot \mathbf{1}^\top$. This term encourages compressible and interpretable solutions without requiring non-quadratic optimization.

\textbf{Temporal Regularization.} To encourage temporal consistency in time-varying sequences, we penalize deviations from the previous frame's solution
$\vw_{\text{prev}}$. This term smooths facial motion and mitigates jitter in the resulting animation.

\subsection{Pose Constraints and Semantic Priors}

In addition to regularization, we enforce several semantic constraints that capture known structure in the blendshape model. These constraints encourage physically or artistically meaningful behavior and prevent implausible or conflicting activations.

\textbf{Bound Constraints.}
All weights are constrained to the interval $[0, 1]$, reflecting their role as interpolation factors between the neutral and full-intensity shapes. 

\textbf{Active Blendshape Subset.}
In many scenarios, only a subset of blendshapes is relevant to a particular character or solve. We allow specifying a set of active blendshapes that restricts all computations, including optimization and constraint evaluation, to the entries of this subset.

\textbf{Canceling Poses (Mutual Exclusion).}
Certain blendshape pairs are semantically incompatible and should not be simultaneously active. For example, the cheek puff and suck expressions represent conflicting muscle actions. Given a list of cancelling pairs of poses, the solver enforces mutual exclusion through a two-pass procedure:
\begin{itemize}
    \item Solve once under $[0, 1]$ bounds.
    \item For each canceling pair, set the upper bound of the smaller activation to 0.
    \item Re-solve with updated bounds.
\end{itemize}

This mechanism is only applied to active pairs, i.e., both shapes are included in the active set.

\textbf{Symmetric Poses.}
Sometimes, blendshapes are defined in symmetric pairs that are expected to activate in coordination. The balanced activation of these symmetric parts is enforced through a regularization term:
\begin{equation}
R_{\text{Sym}}(\vw) = \left\| \mathbf{S} \vw \right\|_2^2,
\end{equation}
where $\mathbf{S} \in \mathbb{R}^{K \times N}$ encodes the pairs of symmetric poses $K$. Each row of $\mathbf{S}$ corresponds to a pair, containing a $+1$ and a $-1$ in the columns of the two poses in the pair and zeros elsewhere. A pair is included only if both poses are active.

\subsection{Data-Guided Blendshape Construction.}

While the solver remains general for any blendshape model, we adopt the ARKit blendshape schema for compatibility with other applications. Canonical ARKit shapes are designed for broad transferability across characters, but they may not precisely capture the expression characteristics of a specific individual. To address this, we refine the ARKit targets to better match the expressions of \audiotoface subjects, balancing generality and subject-specific accuracy.

We construct personalized blendshapes through automatic transfer, targeted personalization, and manual refinement. We first transfer generic ARKit poses from our template onto the neutral mesh of \audiotoface subject. Since our template and subject share consistent topology, directly applying vertex displacements from the template is usually sufficient, though more advanced methods such as deformation transfer~\citep{Sumner2004} can be used when needed. 

We then personalize the blendshape set by replacing certain ARKit poses with expressions from a range of motion (ROM) captures. We also update specific lip shapes using output poses from \audiotoface inference based on a calibration audio designed to cover various ARKit lip shapes. Artists further refine the shapes to ensure clean geometry and precise alignment with the subject's facial structure. 

\section{Results and Use Cases}
\label{sec::results}

In this section, we present the output animations from the Audio2Face-3D networks and demonstrate their application in real-world scenarios.

\cref{fig:results_a2f_identities} illustrates the animation outputs of \audiotoface-v2.3 and \audiotoface-v3.0 networks given the same test audio. Both networks produce natural lip sync animations for the input audio. To create different identities, we utilized the actor-specific v2.3 networks (Mark/Claire/James). For the v3.0 network, we used the same network with different identity conditions.

\cref{fig:results_a2f_emotions} shows the output animations given different emotion conditions. We can generate emotions present in the training data and interpolate them by blending the one-hot emotion vectors. The emotion condition can be keyframed over time, allowing for smooth transitions between emotions. This process can also be automated using the Audio2Emotion network, which detects emotions in audio clips and facilitates transitions for streaming audio. The implicit emotion in the v2.3 networks also modifies the output animation, but in a limited manner. It mainly changes the lip distance or protrusion while maintaining similar motion dynamics.


\newlength{\rowlabelwidth}
\setlength{\rowlabelwidth}{0.12\textwidth}

\begin{figure}[H]
\centering
\setlength{\tabcolsep}{0pt} 
\renewcommand{\arraystretch}{0} 
\begin{tabular}{c c c c c c c c c}
& \emph{\nvGreen{f}oot} & \emph{\nvGreen{w}ear} & \emph{wea\nvGreen{r}} & \emph{cow\nvGreen{b}oy} & \emph{\nvGreen{ch}aps}  & \emph{ch\nvGreen{a}ps}& \emph{j\nvGreen{o}lly} & \emph{\nvGreen{m}oving} \\[0.3em]
\rotatebox{90}{\makebox[0pt][l]{\emph{\small v2.3-Mark}}}\hspace{0.1em} &
\includegraphics[width=\rowlabelwidth]{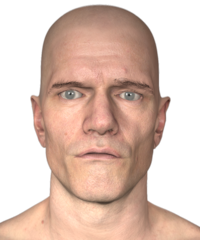} &
\includegraphics[width=\rowlabelwidth]{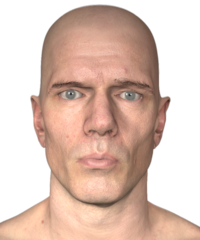} &
\includegraphics[width=\rowlabelwidth]{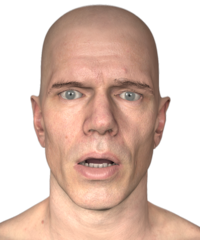} &
\includegraphics[width=\rowlabelwidth]{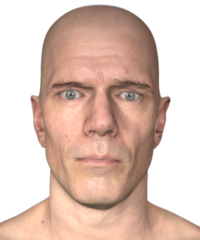} &
\includegraphics[width=\rowlabelwidth]{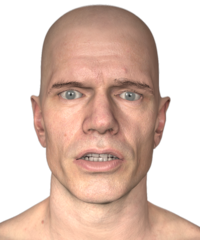} &
\includegraphics[width=\rowlabelwidth]{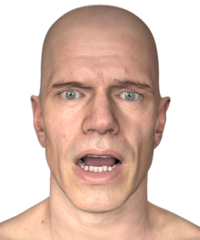} &
\includegraphics[width=\rowlabelwidth]{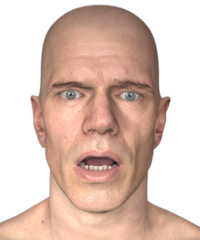} &
\includegraphics[width=\rowlabelwidth]{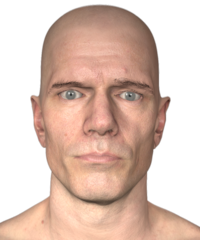} \\[-0.5em]
\rotatebox{90}{\makebox[0pt][l]{\emph{\small v2.3-Claire}}} \hspace{0.1em} &
\includegraphics[width=\rowlabelwidth]{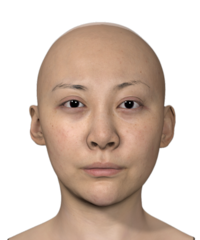} &
\includegraphics[width=\rowlabelwidth]{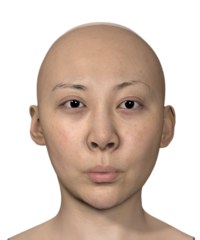} &
\includegraphics[width=\rowlabelwidth]{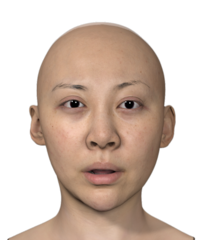} &
\includegraphics[width=\rowlabelwidth]{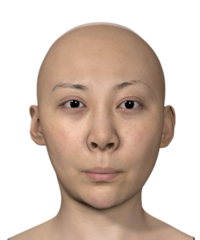} &
\includegraphics[width=\rowlabelwidth]{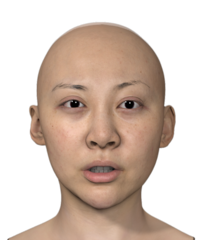} &
\includegraphics[width=\rowlabelwidth]{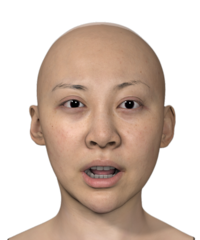} &
\includegraphics[width=\rowlabelwidth]{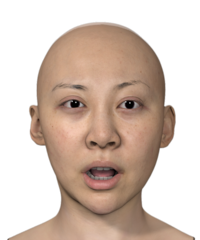} &
\includegraphics[width=\rowlabelwidth]{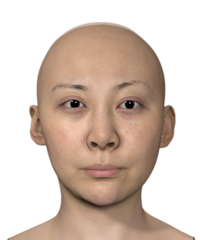} \\
\rotatebox{90}{\makebox[0pt][l]{\emph{\small v2.3-James}}} \hspace{0.1em} &
\includegraphics[width=\rowlabelwidth]{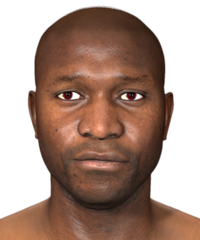} &
\includegraphics[width=\rowlabelwidth]{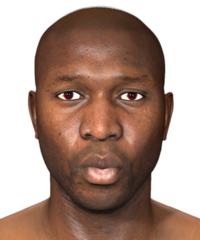} &
\includegraphics[width=\rowlabelwidth]{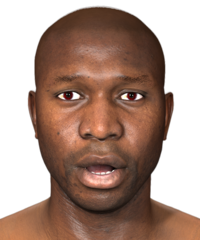} &
\includegraphics[width=\rowlabelwidth]{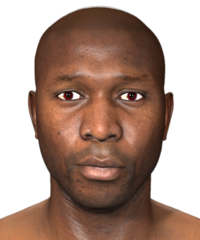} &
\includegraphics[width=\rowlabelwidth]{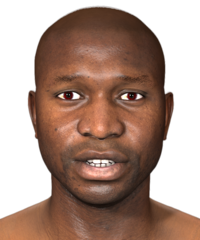} &
\includegraphics[width=\rowlabelwidth]{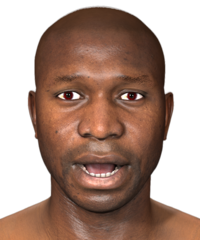} &
\includegraphics[width=\rowlabelwidth]{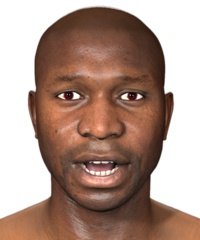} &
\includegraphics[width=\rowlabelwidth]{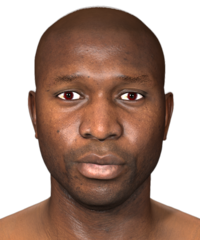} \\
\end{tabular}\\[0.3em]
\textcolor{gray!60}{\rule{\textwidth}{1pt}}\\[0.3em]
\begin{tabular}{c c c c c c c c c}
& \emph{\nvGreen{f}oot} & \emph{\nvGreen{w}ear} & \emph{wea\nvGreen{r}} & \emph{cow\nvGreen{b}oy} & \emph{\nvGreen{ch}aps}  & \emph{ch\nvGreen{a}ps}& \emph{j\nvGreen{o}lly} & \emph{\nvGreen{m}oving} \\[0.3em]
\rotatebox{90}{\makebox[0pt][l]{\emph{\small v3.0 \footnotesize (Mark id) }}} \hspace{0.1em} &
\includegraphics[width=\rowlabelwidth]{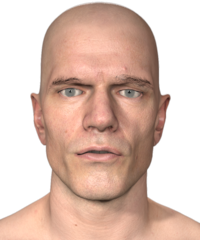} &
\includegraphics[width=\rowlabelwidth]{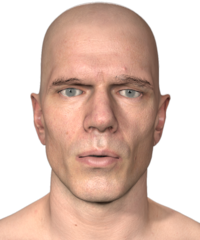} &
\includegraphics[width=\rowlabelwidth]{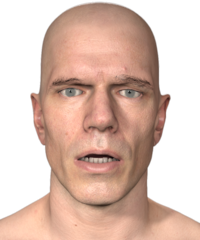} &
\includegraphics[width=\rowlabelwidth]{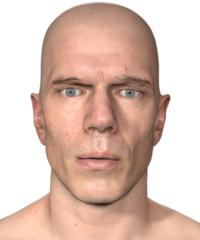} &
\includegraphics[width=\rowlabelwidth]{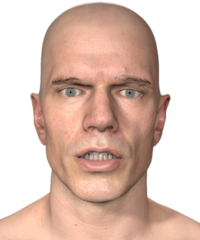} &
\includegraphics[width=\rowlabelwidth]{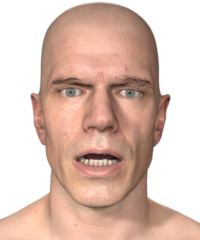} &
\includegraphics[width=\rowlabelwidth]{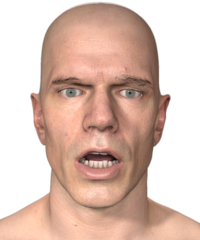} &
\includegraphics[width=\rowlabelwidth]{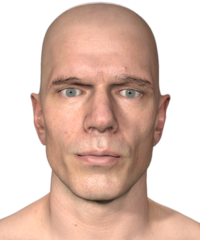} \\[-0.5em]
\rotatebox{90}{\makebox[0pt][l]{\emph{\small v3.0 \footnotesize (Claire id)}}} \hspace{0.1em} &
\includegraphics[width=\rowlabelwidth]{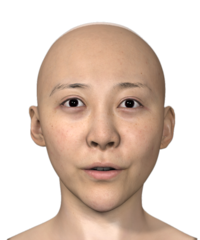} &
\includegraphics[width=\rowlabelwidth]{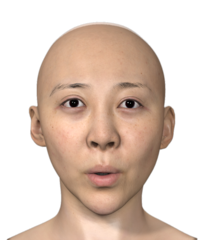} &
\includegraphics[width=\rowlabelwidth]{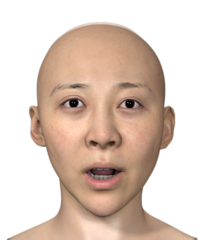} &
\includegraphics[width=\rowlabelwidth]{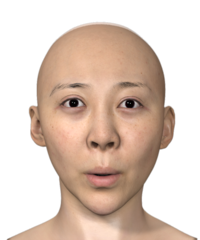} &
\includegraphics[width=\rowlabelwidth]{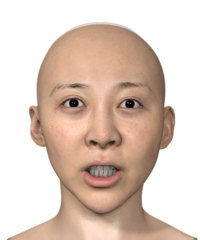} &
\includegraphics[width=\rowlabelwidth]{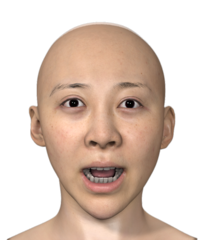} &
\includegraphics[width=\rowlabelwidth]{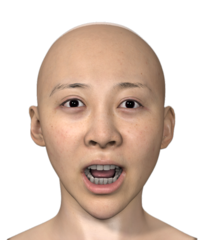} &
\includegraphics[width=\rowlabelwidth]{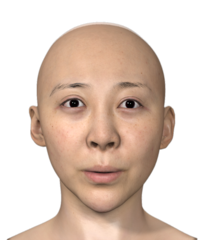} \\
\rotatebox{90}{\makebox[0pt][l]{\emph{\small v3.0 \footnotesize (James id)}}} \hspace{0.1em} &
\includegraphics[width=\rowlabelwidth]{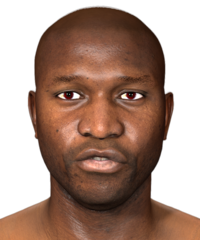} &
\includegraphics[width=\rowlabelwidth]{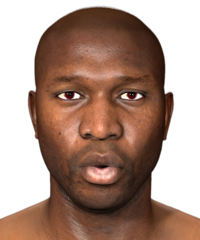} &
\includegraphics[width=\rowlabelwidth]{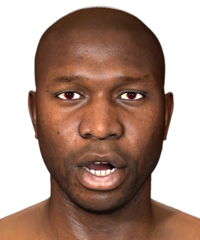} &
\includegraphics[width=\rowlabelwidth]{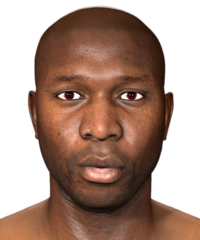} &
\includegraphics[width=\rowlabelwidth]{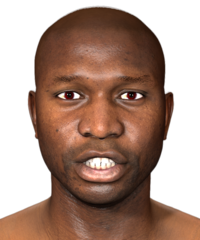} &
\includegraphics[width=\rowlabelwidth]{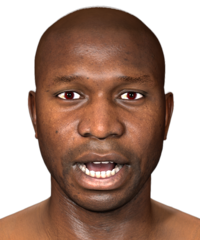} &
\includegraphics[width=\rowlabelwidth]{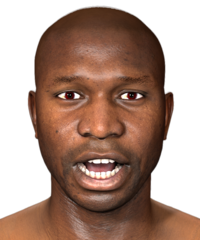} &
\includegraphics[width=\rowlabelwidth]{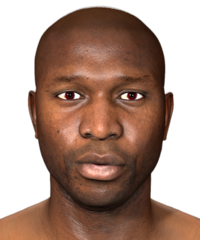} \\
\end{tabular}

\caption{The animation outputs of the \audiotoface-v2.3 (top) and \audiotoface-v3.0 (bottom) networks given the same audio input.}
\label{fig:results_a2f_identities}
\end{figure}
\setlength{\rowlabelwidth}{0.12\textwidth}

\begin{figure}[H]
\centering
\setlength{\tabcolsep}{0pt} 
\renewcommand{\arraystretch}{0.5} 

\vspace{-0.4cm}
\begin{tabular}{c c c c c c c c c}
& \emph{\nvGreen{f}oot} & \emph{\nvGreen{w}ear} & \emph{wea\nvGreen{r}} & \emph{cow\nvGreen{b}oy} & \emph{\nvGreen{ch}aps}  & \emph{ch\nvGreen{a}ps}& \emph{j\nvGreen{o}lly} & \emph{\nvGreen{m}oving} \\[0.3em]
\rotatebox{90}{\makebox[0pt][l]{\emph{\small v3.0 \footnotesize (Mark id)}}} \hspace{0.1em}     &
\includegraphics[width=\rowlabelwidth]{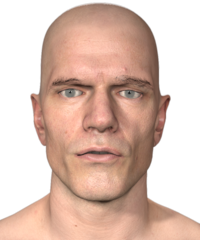}     &
\includegraphics[width=\rowlabelwidth]{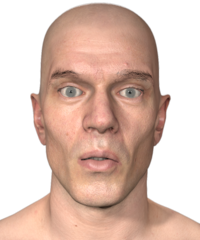}   &
\includegraphics[width=\rowlabelwidth]{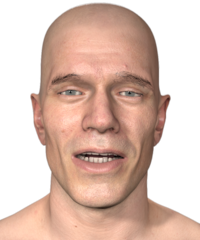}         &
\includegraphics[width=\rowlabelwidth]{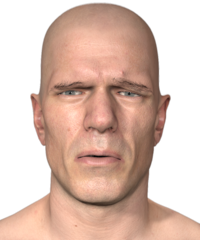}     &
\includegraphics[width=\rowlabelwidth]{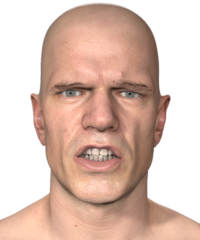}       &
\includegraphics[width=\rowlabelwidth]{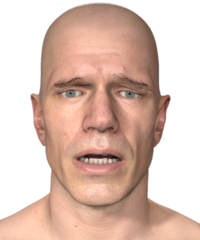}     &
\includegraphics[width=\rowlabelwidth]{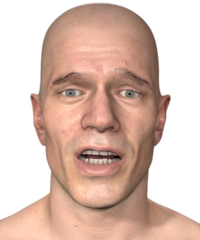}  &
\includegraphics[width=\rowlabelwidth]{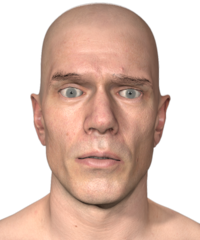}        \\ [-0.75em] 
\rotatebox{90}{\makebox[0pt][l]{\emph{\small v3.0 \footnotesize (Claire id)}}} \hspace{0.1em}   &
\includegraphics[width=\rowlabelwidth]{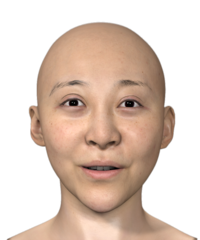}   &
\includegraphics[width=\rowlabelwidth]{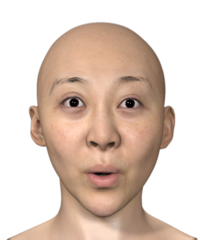} &
\includegraphics[width=\rowlabelwidth]{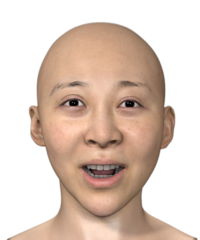}       &
\includegraphics[width=\rowlabelwidth]{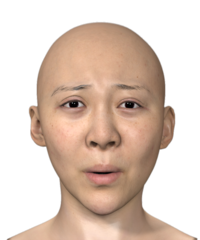}   &
\includegraphics[width=\rowlabelwidth]{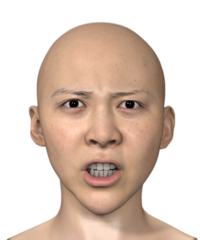}     &
\includegraphics[width=\rowlabelwidth]{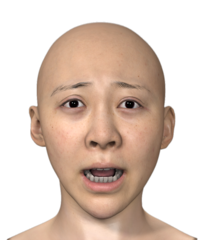}   &
\includegraphics[width=\rowlabelwidth]{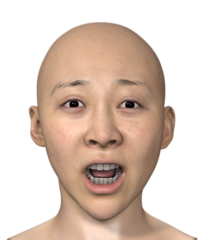}&
\includegraphics[width=\rowlabelwidth]{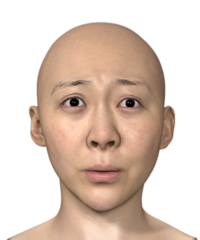}      \\
&\emph{neutral} & \emph{amazement} & \emph{joy} & \emph{disgust} & \emph{anger}  & \emph{sadness}& \emph{cheekiness} & \emph{fear} \\
\end{tabular}\\[0.3em]
\textcolor{gray!60}{\rule{\textwidth}{1pt}}\\[0.3em]

\begin{tabular}{c c c c c c c c c}
& \emph{\nvGreen{f}oot} & \emph{\nvGreen{w}ear} & \emph{wea\nvGreen{r}} & \emph{cow\nvGreen{b}oy} & \emph{\nvGreen{ch}aps}  & \emph{ch\nvGreen{a}ps}& \emph{j\nvGreen{o}lly} & \emph{\nvGreen{m}oving} \\
\rotatebox{90}{\makebox[0pt][l]{\emph{\small v3.0 \footnotesize (James id)}}} \hspace{0.1em} &
\includegraphics[width=\rowlabelwidth]{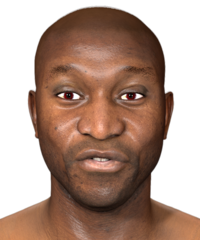} &
\includegraphics[width=\rowlabelwidth]{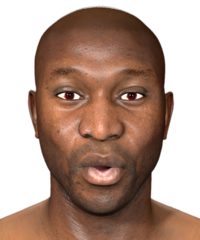} &
\includegraphics[width=\rowlabelwidth]{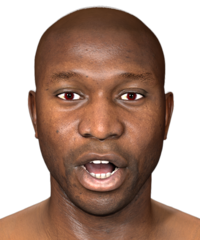} &
\includegraphics[width=\rowlabelwidth]{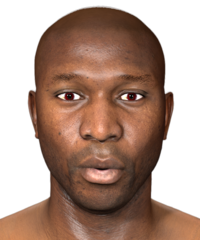} &
\includegraphics[width=\rowlabelwidth]{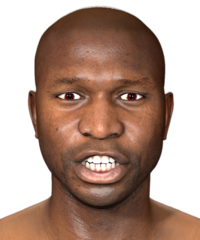} &
\includegraphics[width=\rowlabelwidth]{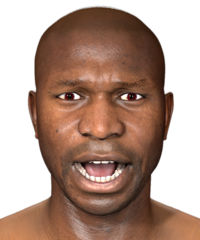} &
\includegraphics[width=\rowlabelwidth]{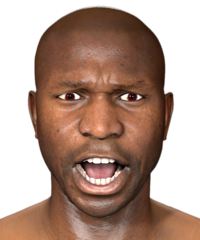} &
\includegraphics[width=\rowlabelwidth]{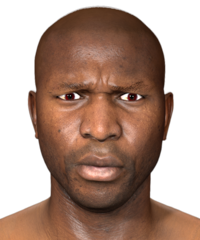} \\[-0.75em] 
\rotatebox{90}{\makebox[0pt][l]{\emph{\small v3.0 \footnotesize (Claire id)}}} \hspace{0.1em}    &
\includegraphics[width=\rowlabelwidth]{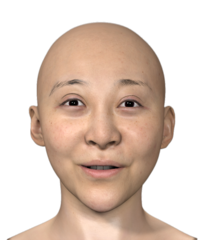} &
\includegraphics[width=\rowlabelwidth]{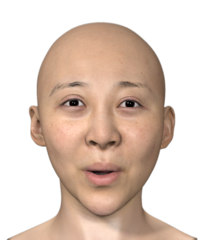} &
\includegraphics[width=\rowlabelwidth]{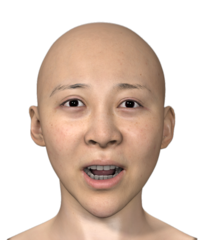} &
\includegraphics[width=\rowlabelwidth]{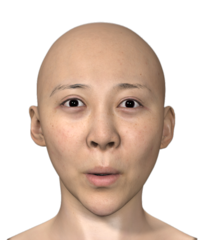} &
\includegraphics[width=\rowlabelwidth]{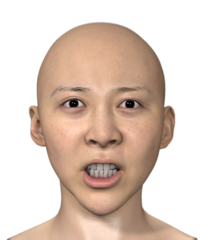} &
\includegraphics[width=\rowlabelwidth]{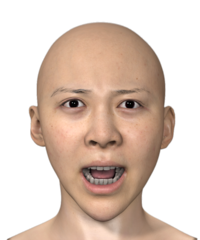} &
\includegraphics[width=\rowlabelwidth]{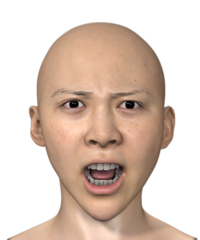} &
\includegraphics[width=\rowlabelwidth]{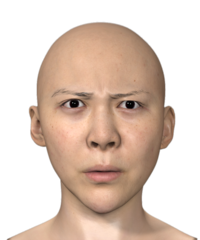} \\[0.3em]
& \makebox[\rowlabelwidth][c]{\shortstack[c]{\emph{\footnotesize joy \scriptsize 1.0} \\\emph{\footnotesize anger \scriptsize 0.0}}}&
\multicolumn{6}{c}{ \raisebox{0.1\rowlabelwidth}{ \tikz[baseline={(current bounding box.center)}]{%
                        \draw[->, thick, color=nvidiagreen, line width=2pt, >=stealth]
                        (0,0) -- (5.90*\rowlabelwidth,0);
                }}} &
\makebox[\rowlabelwidth][c]{\shortstack[c]{\emph{\footnotesize joy \scriptsize 0.0} \\\emph{\footnotesize anger \scriptsize 1.0}}} \\
\end{tabular}
\vspace{-0.1cm}
\caption{Animation results from different emotion conditions. \emph{Top}: Output facial expressions generated from one-hot encoded emotion conditions. \emph{Bottom}: Smooth transition of facial expressions between joy and anger.}
\label{fig:results_a2f_emotions}
\end{figure}


\setlength{\rowlabelwidth}{0.12\textwidth}

\begin{figure}[H]
\vspace{-0.5cm}
\centering
\setlength{\tabcolsep}{0pt}     
\renewcommand{\arraystretch}{0} 
\begin{tabular}{c c c c c c c c c c}
& \emph{\nvGreen{w}ith} & \emph{t\nvGreen{e}nure} & \emph{S\nvGreen{u}zie} & \emph{\nvGreen{a}ll} & \emph{l\nvGreen{ei}sure} & \emph{\nvGreen{f}or} & \emph{y\nvGreen{a}tching}& \emph{\nvGreen{b}ut}\\[0.3em]
\rotatebox{90}{\makebox[0pt][l]{\emph{\small v3.0 \footnotesize (James id)}}} \hspace{0.1em} & 
\includegraphics[width=\rowlabelwidth]{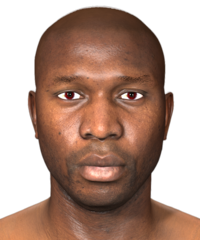} & 
\includegraphics[width=\rowlabelwidth]{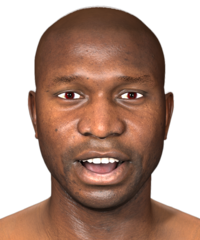} & 
\includegraphics[width=\rowlabelwidth]{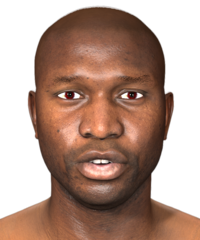} & 
\includegraphics[width=\rowlabelwidth]{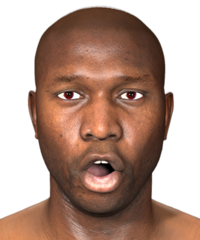} & 
\includegraphics[width=\rowlabelwidth]{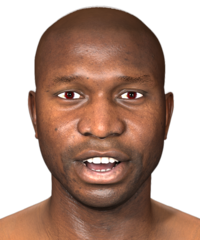} & 
\includegraphics[width=\rowlabelwidth]{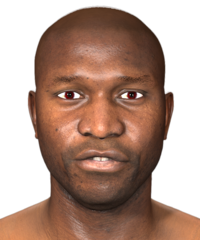} & 
\includegraphics[width=\rowlabelwidth]{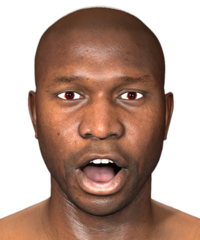} &
\includegraphics[width=\rowlabelwidth]{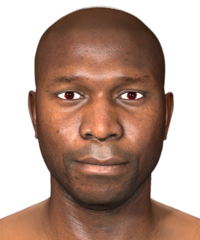} & \\
\rotatebox{90}{\makebox[0pt][l]{\emph{\small Blendshapes}}} \hspace{0.1em} & 
\includegraphics[width=\rowlabelwidth]{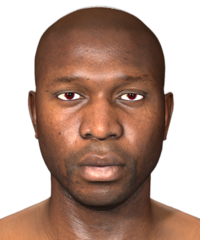} & 
\includegraphics[width=\rowlabelwidth]{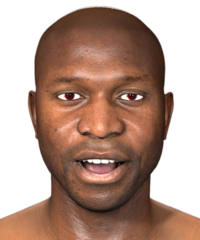} & 
\includegraphics[width=\rowlabelwidth]{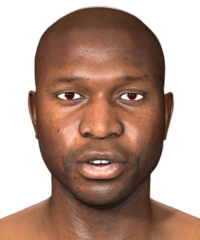} & 
\includegraphics[width=\rowlabelwidth]{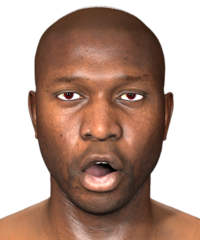} & 
\includegraphics[width=\rowlabelwidth]{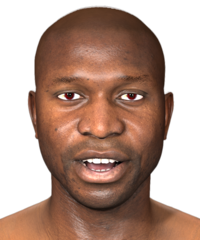} & 
\includegraphics[width=\rowlabelwidth]{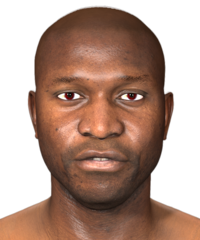} & 
\includegraphics[width=\rowlabelwidth]{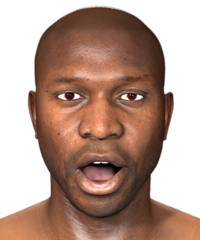} &
\includegraphics[width=\rowlabelwidth]{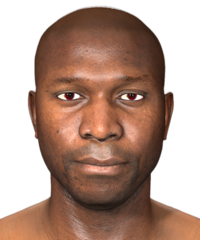} & \\[-0.5em] 
\rotatebox{90}{\makebox[0pt][l]{\emph{\small Retargeting}}} \hspace{0.1em} & 
\includegraphics[width=\rowlabelwidth]{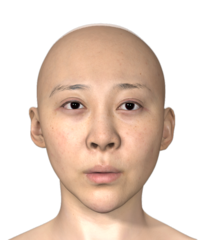} & 
\includegraphics[width=\rowlabelwidth]{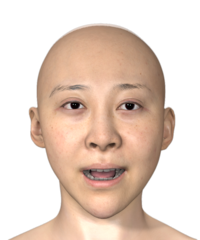} & 
\includegraphics[width=\rowlabelwidth]{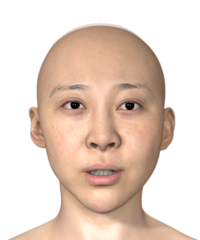} & 
\includegraphics[width=\rowlabelwidth]{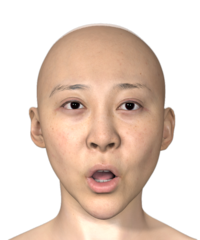} & 
\includegraphics[width=\rowlabelwidth]{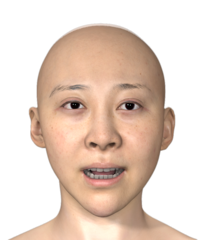} & 
\includegraphics[width=\rowlabelwidth]{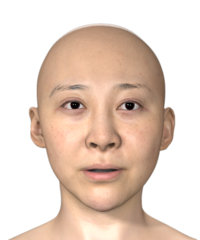} & 
\includegraphics[width=\rowlabelwidth]{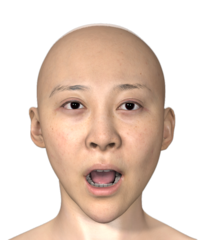} & 
\includegraphics[width=\rowlabelwidth]{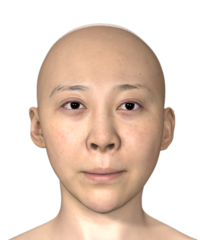} & \\
\rotatebox{90}{\makebox[0pt][l]{\emph{\small Error Magnitude}}} \hspace{0.1em} &
\includegraphics[width=\rowlabelwidth]{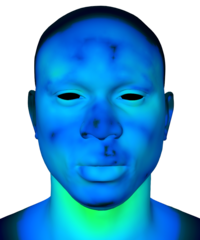} & 
\includegraphics[width=\rowlabelwidth]{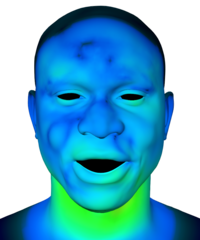} & 
\includegraphics[width=\rowlabelwidth]{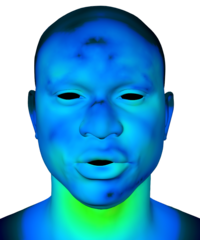} & 
\includegraphics[width=\rowlabelwidth]{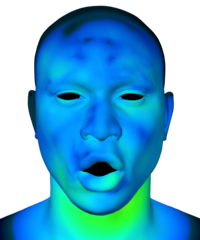} & 
\includegraphics[width=\rowlabelwidth]{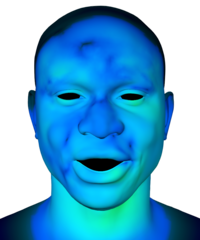} & 
\includegraphics[width=\rowlabelwidth]{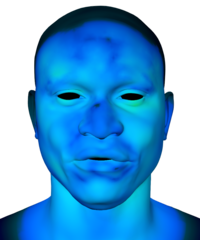} & 
\includegraphics[width=\rowlabelwidth]{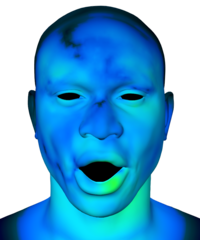} &
\includegraphics[width=\rowlabelwidth]{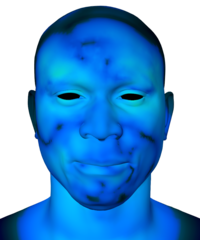}&
\hspace{-0.5em}
\begin{tikzpicture}
\begin{axis}[
    ymin=0, ymax=8,
    hide axis,
    scale only axis,
    height=0.13\textwidth,
    colorbar,
    colormap/jet,
    point meta min=0,
    point meta max=8,
    axis line style= {white},
    colorbar style={
        draw=none,
        width =0.007\textwidth, 
        height=0.13\textwidth,
        ytick={0,4,8},
        yticklabels={0 mm, 4 mm, 8 mm},
        tick align=center,
        tick label style={font=\tiny}
    }
]
\addplot [draw=none] coordinates {(0,0) (0.,8)};
\end{axis}
\end{tikzpicture}
\end{tabular}
\caption{Inferred animation, blendshape reconstruction, and retargeted animation on a new identity with equivalent blendshapes. 
}
\label{fig:results_bs_solve}
\end{figure}
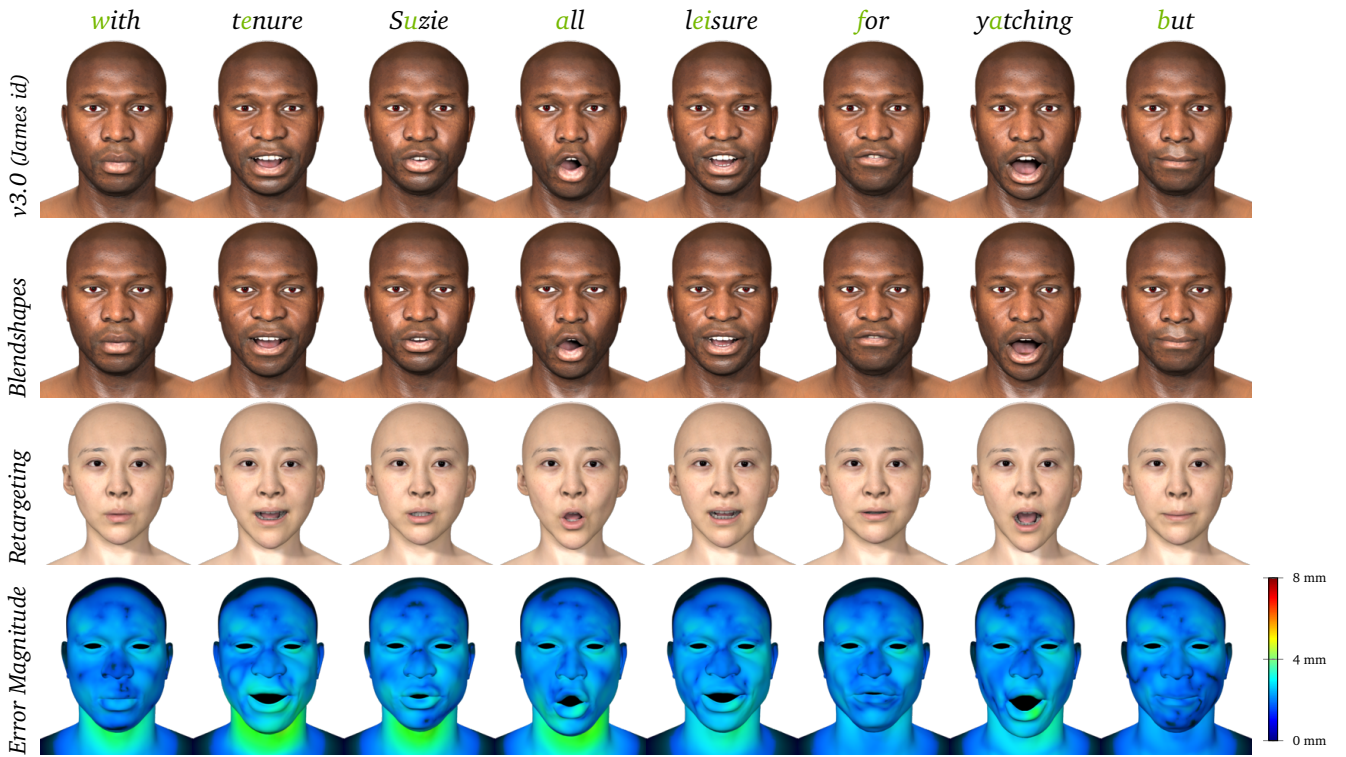

The blendshape solve converts the output animation into a blendshape-based animation. \cref{fig:results_bs_solve} compares the raw output animation with the converted blendshape-based animation. In our open-source version, we achieve blendshape-based animations that closely match the original output, enabling retargeting to new characters while preserving the same motion semantics.

\textbf{Computation Time and Memory Consumption.} \cref{tab:model_acceleration} shows the inference FPS of \audiotoface networks with different inference engines on a GeForce RTX 4090 GPU. We achieve faster inference speeds and support multiple concurrent processes on a single GPU by converting the networks to TensorRT formats. The TensorRT inference achieves faster inference speeds and maintains high per-track FPS during multi-track inference, as shown in \cref{tab:model_acceleration}. The subsequent blendshape solve adds negligible computation time. Specifically, computing the blendshape weights for skin and tongue takes 0.3 ms and scales linearly with the number of concurrent processes. 

GPU memory consumption during inference is one of the critical factors in using deep-learning solutions. \audiotoface-v2.3 network in TensorRT uses 0.6 GB of GPU memory to run 8 concurrent tracks and scales well to a large number of tracks (e.g., 1.3 GB for 128 tracks). \audiotoface-v3.0 network in TensorRT consumes 1.4 GB and 4.0 GB of GPU memory to run 1 and 8 concurrent tracks, respectively. The difference in memory usage is primarily due to the different output sizes from a single inference: \audiotoface-v2.3 outputs 1-frame data in a PCA-compressed format, while \audiotoface-v3.0 outputs 30-frame data in a raw vertex format.

\begin{table}[ht]
    \setlength{\tabcolsep}{4.7pt}
    \small
    \captionsetup{justification=centering}
    \caption{Inference FPS with different inference engines on a single GeForce RTX 4090. The numbers next to the inference engines indicate the number of concurrent inference tracks.}
    \centering
    \setlength{\tabcolsep}{6pt}
    \begin{tabularx}{\textwidth}{>{\raggedleft\arraybackslash}X>{\centering\arraybackslash}X>{\centering\arraybackslash}X>{\centering\arraybackslash}X>{\centering\arraybackslash}X>{\centering\arraybackslash}X>{\centering\arraybackslash}X}
     \toprule
     Inference Engine & PyTorch (1) & ONNX (1) & TensorRT (1) & TensorRT (2) & TensorRT (4) & TensorRT (8) \\ 
     \toprule
     v2.3-Claire  & 194 & 252 & 453 & 451 & 433 & 413 \\
     \midrule
     v3.0  & 2069 & 1803 & 3269 & 2818 & 2083 & 1250 \\
     \bottomrule
    \end{tabularx}
    
    \label{tab:model_acceleration}
\end{table}

\textbf{Quality Benchmark.} To evaluate the performance and quality of the \audiotoface system, we have established a comprehensive quality benchmark. This benchmark includes multiple metrics designed to assess various quality aspects of the generated facial animations. Specifically, we employ SyncNet, the jitter metric, the bilabial sound score, and the expressiveness score as our quality benchmarks:

\begin{itemize}
    \item To evaluate lip sync accuracy, we use the pre-trained 3D \emph{SyncNet} from GeneFace~\citep{ye2023geneface}, which was trained on the large-scale LRS3 dataset using 3D landmark and audio data to discriminate synchronized versus out-of-sync lip landmark-audio pairs. For evaluation, we generate animation cache from the test talking audio, extract mouth-region landmark animation sequences and HuBERT audio features. For each input pair of mouth landmark motion sequence and audio feature, the model outputs a sync score in [0,1], where higher scores indicate better lip sync. 
    \item The \emph{jitter metric} quantifies the smoothness of facial animations by measuring the Fourier metric and Fr\'echet distance in facial motion. The
    \emph{Fourier metric} quantifies the intensity of the high-frequency component of facial animation around the mouth region. To compute the metric, we normalize the vertex acceleration, extract the mouth region, and apply a Fast Fourier Transform (FFT) to convert the data to the frequency domain. The high-frequency intensity is calculated based on an experimental cutoff of the frequency magnitudes, with lower intensity indicating natural-looking animation with less jitter.
    The \emph{Fr\'echet distance} metric calculates the Fr\'echet distance between ground truth (GT) and predicted vertex animations, focusing on the mouth region. This metric provides a comprehensive spatial and temporal evaluation of the predicted animation quality by comparing it to the GT. A lower Fr\'echet distance indicates a closer match to the GT and less jitter in the predicted animation.
    \item The \emph{bilabial sound score} quantifies the precision of lip closure during the articulation of bilabial phonemes /M/, /B/, /P/, which require precise lip contact for correct speech production. It is calculated by aligning audio to phoneme sequences using predictive aligners~\citep{zhu2022Charsiu} and extracting the time intervals of the target bilabials. For each instance, the vertical distance between predefined upper and lower lip vertices is measured frame-by-frame, and the minimum distance during the duration of the phoneme determines the success. A phoneme is marked as successful if this distance falls below a defined threshold. The final score reports the percentage of successful closures in all bilabial instances.
    \item The \emph{expressiveness score} quantifies the emotional intensity of inferred facial expressions. For each emotion sequence, the system computes frame-by-frame neutralized FACEM features (FACial Expression Measurement)~\citep{Katsikitis2003} normalized by feature-specific ranges to account for anatomical differences. Scores are obtained by averaging the normalized features over time, reflecting the intensity of deviation of an expression from neutrality. The system also compares each emotion to other emotional references to assess pairwise expressivity consistency. This approach enables robust comparisons of emotional expressiveness across different characters and model versions and aligns well with human perception of emotional intensity.
\end{itemize}

We use this benchmark primarily for sanity checking and as a reference; it is not critical in choosing the best network. The main evaluation of network quality is conducted through subjective assessments by experts, who provide nuanced feedback on the realism and effectiveness of the generated animations.

\textbf{Limitations.} While \audiotoface is a powerful solution for realistic facial animation, the system has certain limitations. If the audio has severe background noise, strong non-verbal sounds, or requires unique lip shapes beyond the training data, the system may fail to generate accurate lip movements. Additionally, the system may struggle with generating semantically meaningful upper-face or eyeball motion. Humans often alter their expressions based on the conversation context, which requires an understanding of the semantics. Since \audiotoface processes audio data from a short temporal window, the resulting upper-face and eyeball animations appear simple and fail to capture the semantic context. Lastly, the system cannot generate natural idle motion or listening motion from long silent audio or speech audio from a conversation partner, which are important for natural avatar interactions.

\subsection{\audiotoface Use Cases}
\label{subsec:use_case}

\begin{figure}[H]
    \centering
    \includegraphics[width=\textwidth]{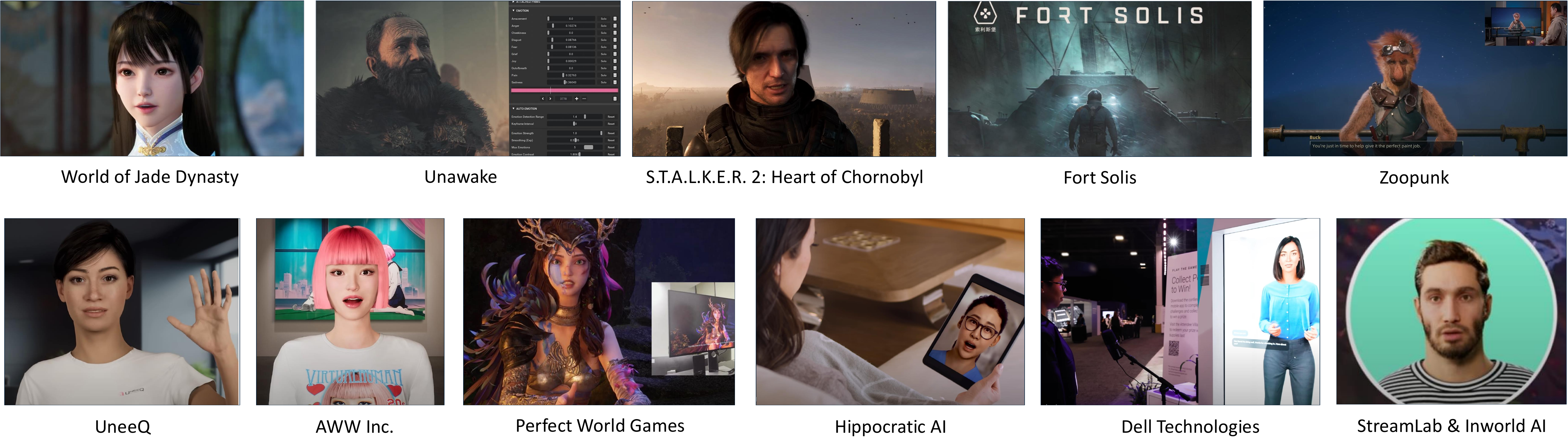}
    \caption{Example use cases of \audiotoface in games (top row) and interactive avatars (bottom row).}
    \label{fig:results_use_cases}
\end{figure}

\audiotoface is particularly efficient in two domains: gaming and AI-driven interactive avatars. Games often require hundreds of hours of facial animation, which can be more efficiently produced through audio-driven animation compared to video capture. In interactive AI-avatar applications, avatars typically receive only audio input from large language models (LLM) and text-to-speech (TTS) systems, necessitating the generation of corresponding facial animations at an interactive rate for a natural conversation. \cref{fig:results_use_cases} shows our real-world use cases in the game and interactive avatar domain. Our customers could either generate a large amount of facial animation automatically using \audiotoface and set up an interactive avatar for various domains such as customer service and gaming characters.

\section{Related Work}
\label{sec::related_work}
In recent years, substantial progress has been achieved in audio-driven facial animation, including both 2D talking head and 3D animation generation.
\subsection{Audio-Driven 2D Talking Head Generation}
Recent advancements in 2D talking head generation have focused on achieving precise lip-sync and natural head movements, leveraging various deep learning techniques such as neural networks, 3D neural rendering, and diffusion models.
As an early approach, Wav2Lip~\citep{Prajwal2020Wav2Lip} focuses on precise lip-sync for arbitrary identities in dynamic videos by employing a lip-sync discriminator.
Subsequent methods like MakeItTalk~\citep{Zhou2020MakeltTalk} and Audio2Head~\citep{wang2021audio2head} generate talking-head videos from a single image by not only achieving accurate lip-sync but also modeling natural head motions.

Another stream leverages 3D neural rendering techniques, which enable the generation of high-quality 2D rendered videos by learning 3D information from multiple viewpoints.
For instance, AD-NeRF~\citep{guo2021adnerf} and SD-NeRF~\citep{Shen2024sdnerf} leverage Neural Radiance Fields (NeRF)~\citep{Mildenhall2021NeRF}, which effectively models delicate face features with multi-view consistency.
Additionally, GaussianTalker~\citep{cho2024gaussiantalker} leverages an efficient rendering capability of 3D Gaussian Splatting~\citep{Kerbl2023Gaussian} for real-time pose-controllable talking heads, encoding 3D Gaussian attributes into a shared implicit feature representation merged with audio features.

With the advent of diffusion models~\citep{ho2020denoising, song2020denoising,rombach2022stable}, recent methods have successfully integrated diffusion architecture into audio-driven 2D talking head generation, achieving impressive performance. 
DiffTalk~\citep{shen2023difftalk} models talking head generation as an audio-driven denoising process, incorporating reference face images and landmarks to generalize well across different identities.
AniTalker~\citep{Liu2024AniTalker} integrates a diffusion model to capture subtle expressions and head movements using universal motion representation and self-supervised learning strategies, while EchoMimic~\citep{chen2025echomimic} combines audio-driven and landmark-driven methods with a diffusion model.

Further enhancing expressiveness and personalization, recent approaches have enabled greater control over emotion or style. 
GC-AVT~\citep{Liang2022GCAVT} achieves precise control of lip movements, head poses, and facial expressions by decoupling audio-visual sources. 
StyleTalk~\citep{Ma2023StyleTalk} introduces a framework for generating talking heads with diverse speaking styles using a style encoder and a style-aware adaptive transformer. 
TalkCLIP~\citep{Yifeng2023talkclip} allows specifying facial expressions via natural language, leveraging a style encoder based on CLIP~\citep{radford2021learning} to map text descriptions to expression representations.

\begin{table}[t]
\centering
\caption{Comparison of different audio-driven 3D facial animation generation models and their capabilities. Ours-Reg and Ours-Diff indicates regression-based network (\audiotoface-v2.3) and diffusion-based network (\audiotoface-v3.0), respectively.}
\label{tab:related-work-comparison}
\begin{tabular}{lccccc}
\toprule
Model & Base Network & Animating Area &  Emotion & Identity & Inference
\\
\midrule
Taylor et al. & MLP & Lower face & \textcolor{red}{\xmark} & Retargeting & Real-time streaming \\
Karras et al.  & CNN & Full face & \nvGreen{\checkmark} & Single & Real-time streaming \\
Visemenet et al.  & LSTM & Lower face & \textcolor{red}{\xmark} & Single & Real-time \\
VOCA     & CNN & Full face & \textcolor{red}{\xmark} & Multi & Offline \\
MeshTalk   & CNN  & Full face & \nvGreen{\checkmark} & Single & Offline \\
FaceFormer  & Transformer & Full face & \textcolor{red}{\xmark} & Multi & Offline \\
FaceXHuBERT  & Transformer+GRU & Full face & \nvGreen{\checkmark} & Multi & Offline \\
FaceDiffuser & Diffusion & Full face & \textcolor{red}{\xmark} & Multi & Offline \\
FaceTalk      & Diffusion & Full face & \nvGreen{\checkmark} & Multi & Offline \\
DiffPoseTalk   & Diffusion & Full face+head & \textcolor{red}{\xmark} & Multi & Offline \\
\midrule
\textbf{Ours-Reg} & CNN & Full face & \nvGreen{\checkmark} & Single & Real-time streaming \\ 
\textbf{Ours-Diff} & Diffusion & Full face & \nvGreen{\checkmark} & Multi & Real-time Streaming \\ 
\bottomrule
\end{tabular}
\end{table}

\subsection{Audio-Driven 3D Facial Animation Generation}
The early approaches in audio-driven 3D facial animation focus on lip movements.
For example, Taylor et al.~\citep{Taylor2017deep} present a deep learning method with a sliding window predictor that focuses on mapping phoneme labels from the audio to mouth movements. 
Similarly, Visemenet~\citep{Zhou2018Visemenet} proposes a three-stage LSTM network that produces animator-centric speech motion curves, ensuring real-time lip synchronization from audio.
Building on these efforts, Karras et el.~\citep{karras2017afa} introduce an end-to-end CNN-based method that maps audio to 3D vertex coordinates in real-time and suggest a way to extract emotion controls from the data.
VOCA~\citep{Cudeiro2019VOCA} also presents a neural network model that generalizes well to unseen subjects, offering controls for speaking style and facial movements, introducing a unique 4D face dataset.
MeshTalk~\citep{Richard2021MeshTalk} advances the field by addressing upper face animation and co-articulation with a categorical latent space that disentangles audio-correlated and uncorrelated information, leading to highly realistic full-face animations. 
To enhance generalization to longer audio sequences, FaceFormer~\citep{Fan2022FaceFormer} leverages a Transformer-based autoregressive model and self-supervised pre-trained speech representations.

Research has also focused on incorporating emotion or style control and enhancing animation quality and accuracy. 
FaceXHuBERT~\citep{haque2023facexhubert}, which is based on the self-supervised HuBERT model, employs a binary emotion label to control the expressivity of the generated animation. 
ExpCLIP~\citep{Zhong2024ExpCLIP}, MMHead~\citep{Wu2024MMHead}, and  Jung et al.~\citep{Jung2024Audio} leverage a language model such as CLIP~\citep{radford2021learning} or ChatGPT~\citep{openai2023chatgpt} to enable text-induced emotion control.
More recently, diffusion models have been utilized to enhance the realism of synthesized animations.
FaceDiffuser~\citep{stan2023facediffuser} addresses the non-deterministic nature of non-verbal facial cues by training a diffusion-based model with both 3D vertex and blendshape datasets.
FaceTalk~\citep{Aneja2024FaceTalk} employs a latent diffusion model~\citep{rombach2022stable} in the expression space of neural parametric head models to synthesize head motion sequences from audio signals. 
DiffPoseTalk~\citep{Sun2024DiffPoseTalk} introduces a diffusion-based generative framework that combines speech input with style embeddings extracted from short reference videos, enabling stylistic generation.

In our approach, we utilize CNN and diffusion backbones to synthesize high-quality audio-driven 3D facial animations. 
Our method supports real-time streaming inference with emotion control, and the diffusion-based network also accommodates multiple identities. 
Table~\ref{tab:related-work-comparison} provides a comparison of audio-driven 3D face animation generation models and their capabilities.


\section{Experimental Features}
\label{sec::future}
In addition to the core \audiotoface features in the current open-source distribution, we have experimented with the following additional features that effectively extend \audiotoface. In this section, we describe these experimental features and provide a preview of their results.

\subsection{Text-driven Emotion and Facial Movement Control}
Describing emotional states or facial expressions using natural language offers significant advantages over simply using pre-defined sliders. We enable text-driven emotion and facial movements control in the diffusion-based network by leveraging a CLIP text encoder~\citep{radford2021learning}.
We simply replace the one-hot emotion vector $\ve$ in Fig~\ref{fig:diffusion_network} with a CLIP text embedding $\vc$.

\textbf{Training Data.} We use the same 11 different emotional states as described in Sec.~\ref{subsec:capture}. For each state, we annotate various emotion description prompts with GPT-4o~\citep{OpenAI2023GPT4}
(e.g., ``appears neutral'' and ``looks indifferent'' for neutral emotion, ``is angry'' and ``expresses furiously'' for angry emotion, etc.).
Also, we manually translate ARKit face blendshape attributes into short facial movement descriptions (e.g., ``BrowInnerUp'' to ``raised inner eyebrows'').
Following previous work~\citep{Yifeng2023talkclip}, we create the prompts describing emotion and facial movements with a predefined format: ``a person \{emotion\} and speaks with \{facial movement\}''.
For more detailed control, we include adverbs to describe the intensity of emotions or facial movements.
We annotate the emotion intensity by calculating the expressiveness score (see Sec.~\ref{sec::results}) with neutral pose.
That is, the further from the neutral pose, the higher the emotion intensity.
Additionally, we utilize the weights of the ARKit face blendshape attributes to determine facial movement intensity. 
The \textit{intensity adverb} is added to the prompt only when the intensity score exceeds a predefined threshold.
Accordingly, examples of the final prompt are ``a person looks joyful and speaks with raised eyebrows and smiling mouth'' or ``a person is \textit{greatly} breathless and speaks with \textit{immensely} widened eyes.''.

\textbf{Emotion Transition.} We can also model smooth emotion transitions by feeding different emotions to each frame. 
We can achieve this by interpolating two different CLIP text embeddings $\vc_1$ and $\vc_2$ as follows:

\begin{equation}
\vc = \alpha \vc_1 + (1 - \alpha) \vc_2,
\end{equation}
where $\alpha$ gradually decreases from 1 to 0 when performing the emotion transition from $\vc_1$ to $\vc_2$.
We found that directly changing $\alpha$ from 1 to 0 also works well.
This is because the GRU-based decoder, which references the previous output when generating the current one, effectively supports a smooth transition between emotions.

\textbf{Results.} Fig.~\ref{fig:results_text_cond} illustrates that our model generates output with various emotions and facial movements.
(a) presents text-driven emotion control, including the unseen emotion `lonely'. 
Thanks to the prior knowledge of the CLIP text encoder, the model can be generalized to emotions that were unseen during training.
Additionally, (b) and (c) demonstrate the capability of controlling facial movement and emotion intensity using text.
Lastly, (d) shows a smooth transition of emotion from joy to anger using interpolation between the CLIP text embeddings.
The results are generated with a test audio.




\setlength{\rowlabelwidth}{0.12\textwidth}

\begin{figure}[H]
\definecolor{PromptGray}{gray}{0.6}

\setlength{\tabcolsep}{0pt} 
\renewcommand{\arraystretch}{0.5} 

\newlength{\hgap}
\setlength{\hgap}{-1.25em} 

\newcommand{\xdownarrow}[1]{\mathrel{\rotatebox{-90}{$\xrightarrow{#1}$}}}
\newcommand{\promptText}[1]{\textcolor{PromptGray}{\small \emph{#1}}}


\begin{flushleft}
(a) Emotion Control
\end{flushleft}
\centering
\begin{tabular}{m{3.1cm} c c c c c c c c }

& \emph{t\nvGreen{ou}rism}    & \hspace{\hgap}
\emph{con\nvGreen{t}ribution} & \hspace{\hgap} 
\emph{\nvGreen{f}our}         & \hspace{\hgap} 
\emph{t\nvGreen{we}nty}       & \hspace{\hgap} 
\emph{\nvGreen{se}ven}        & \hspace{\hgap} 
\emph{bh\nvGreen{a}t}         & \hspace{\hgap} 
\emph{\nvGreen{b}illion}      & \hspace{\hgap} 
\emph{t\nvGreen{e}n} \\[-0.2em]

\promptText{``a person looks joyful''} &
\CenterImage{\includegraphics[width=\rowlabelwidth]{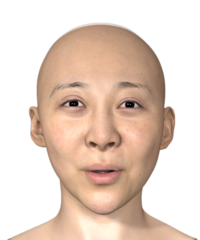}} & \hspace{\hgap}
\CenterImage{\includegraphics[width=\rowlabelwidth]{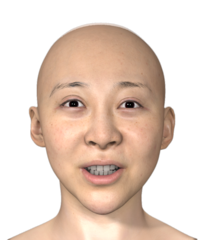}} & \hspace{\hgap}
\CenterImage{\includegraphics[width=\rowlabelwidth]{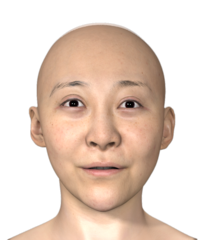}} & \hspace{\hgap}
\CenterImage{\includegraphics[width=\rowlabelwidth]{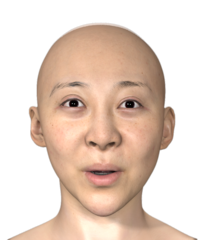}} & \hspace{\hgap}
\CenterImage{\includegraphics[width=\rowlabelwidth]{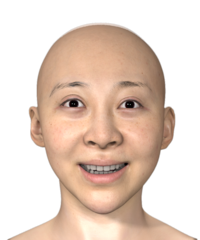}} & \hspace{\hgap}
\CenterImage{\includegraphics[width=\rowlabelwidth]{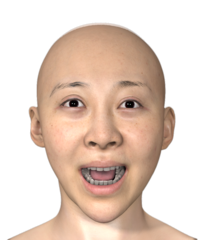}} & \hspace{\hgap}
\CenterImage{\includegraphics[width=\rowlabelwidth]{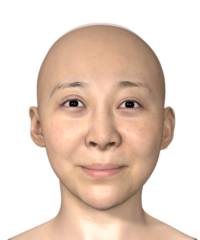}} & \hspace{\hgap}
\CenterImage{\includegraphics[width=\rowlabelwidth]{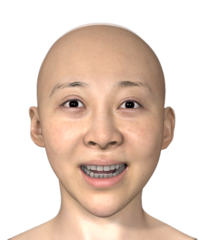}} \\ 
\promptText{``a person is amazed''} & 
\CenterImage{\includegraphics[width=\rowlabelwidth]{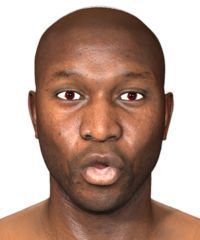}} & \hspace{\hgap}
\CenterImage{\includegraphics[width=\rowlabelwidth]{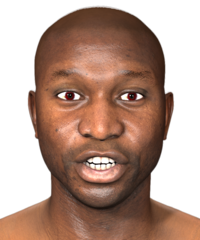}} & \hspace{\hgap}
\CenterImage{\includegraphics[width=\rowlabelwidth]{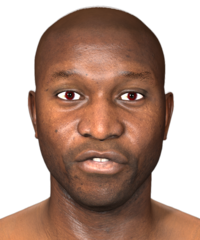}} & \hspace{\hgap}
\CenterImage{\includegraphics[width=\rowlabelwidth]{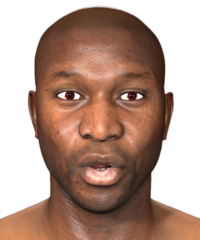}} & \hspace{\hgap}
\CenterImage{\includegraphics[width=\rowlabelwidth]{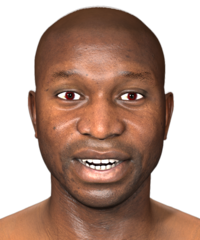}} & \hspace{\hgap}
\CenterImage{\includegraphics[width=\rowlabelwidth]{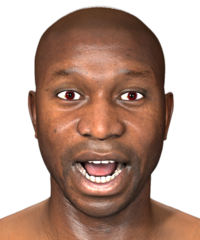}} & \hspace{\hgap}
\CenterImage{\includegraphics[width=\rowlabelwidth]{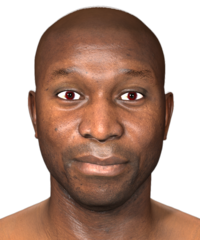}} & \hspace{\hgap}
\CenterImage{\includegraphics[width=\rowlabelwidth]{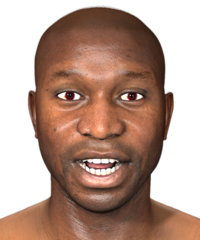}} \\ 
\promptText{``a person is disgusted''} & 
\CenterImage{\includegraphics[width=\rowlabelwidth]{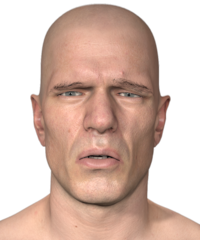}} & \hspace{\hgap}
\CenterImage{\includegraphics[width=\rowlabelwidth]{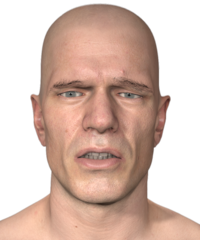}} & \hspace{\hgap}
\CenterImage{\includegraphics[width=\rowlabelwidth]{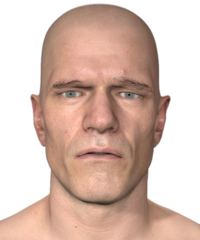}} & \hspace{\hgap}
\CenterImage{\includegraphics[width=\rowlabelwidth]{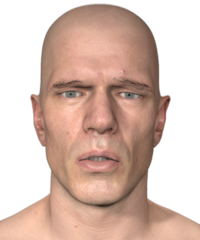}} & \hspace{\hgap}
\CenterImage{\includegraphics[width=\rowlabelwidth]{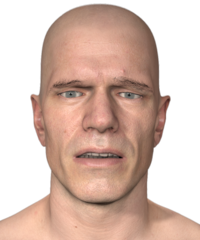}} & \hspace{\hgap}
\CenterImage{\includegraphics[width=\rowlabelwidth]{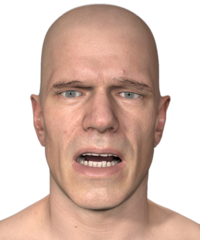}} & \hspace{\hgap}
\CenterImage{\includegraphics[width=\rowlabelwidth]{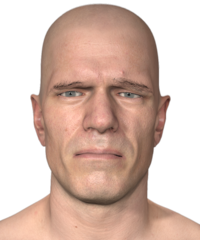}} & \hspace{\hgap}
\CenterImage{\includegraphics[width=\rowlabelwidth]{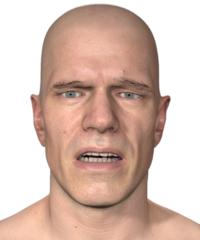}} \\ [-0.5em]
\promptText{``a person is lonely''}  & 
\CenterImage{\includegraphics[width=\rowlabelwidth]{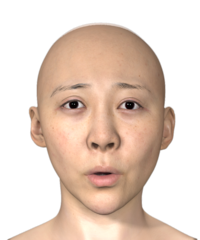}} & \hspace{\hgap}
\CenterImage{\includegraphics[width=\rowlabelwidth]{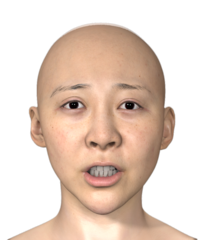}} & \hspace{\hgap}
\CenterImage{\includegraphics[width=\rowlabelwidth]{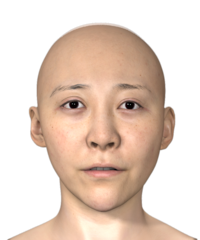}} & \hspace{\hgap}
\CenterImage{\includegraphics[width=\rowlabelwidth]{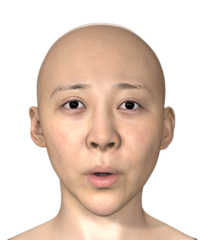}} & \hspace{\hgap}
\CenterImage{\includegraphics[width=\rowlabelwidth]{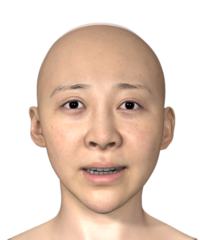}} & \hspace{\hgap}
\CenterImage{\includegraphics[width=\rowlabelwidth]{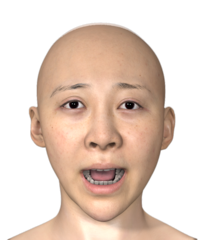}} & \hspace{\hgap}
\CenterImage{\includegraphics[width=\rowlabelwidth]{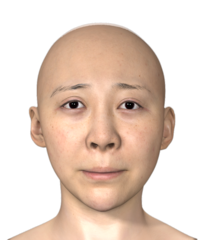}} & \hspace{\hgap}
\CenterImage{\includegraphics[width=\rowlabelwidth]{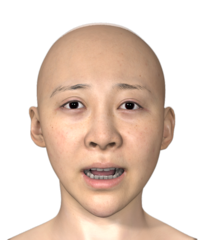}} \\
\end{tabular}\\[0.2em]
\textcolor{gray!60}{\rule{\textwidth}{1pt}}\\[0.2em]

\begin{flushleft}
(b) Facial Movement Control
\end{flushleft}
\begin{tabular}{m{3.1cm} c c c c c c c c }
\parbox{2.9cm}{\promptText{``a person appears neutral and speaks with smiling mouth''}} &
\CenterImage{\includegraphics[width=\rowlabelwidth]{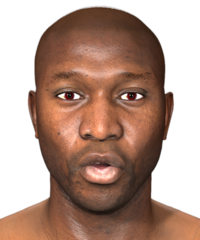}} & \hspace{\hgap}
\CenterImage{\includegraphics[width=\rowlabelwidth]{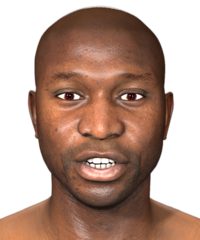}} & \hspace{\hgap}
\CenterImage{\includegraphics[width=\rowlabelwidth]{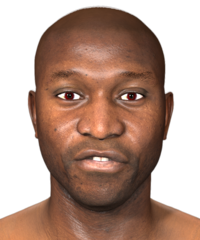}} & \hspace{\hgap}
\CenterImage{\includegraphics[width=\rowlabelwidth]{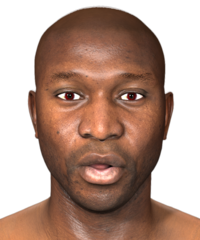}} & \hspace{\hgap}
\CenterImage{\includegraphics[width=\rowlabelwidth]{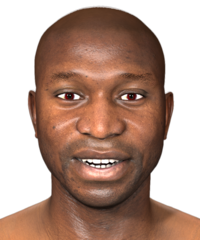}} & \hspace{\hgap}
\CenterImage{\includegraphics[width=\rowlabelwidth]{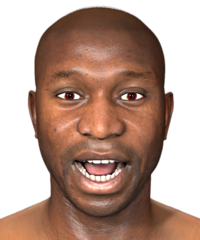}} & \hspace{\hgap}
\CenterImage{\includegraphics[width=\rowlabelwidth]{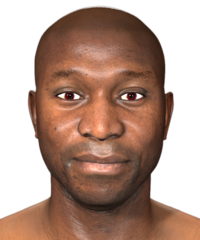}} & \hspace{\hgap}
\CenterImage{\includegraphics[width=\rowlabelwidth]{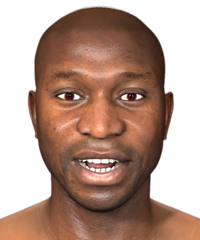}} \\
\parbox{3cm}{\centering\promptText{``a person appears neutral and speaks with furrowed eyebrows''}} &
\CenterImage{\includegraphics[width=\rowlabelwidth]{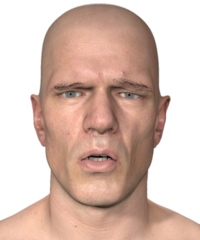}} & \hspace{\hgap}
\CenterImage{\includegraphics[width=\rowlabelwidth]{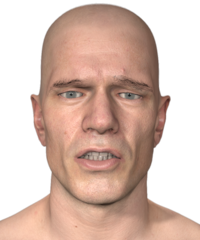}} & \hspace{\hgap}
\CenterImage{\includegraphics[width=\rowlabelwidth]{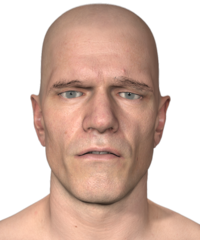}} & \hspace{\hgap}
\CenterImage{\includegraphics[width=\rowlabelwidth]{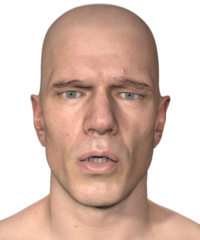}} & \hspace{\hgap}
\CenterImage{\includegraphics[width=\rowlabelwidth]{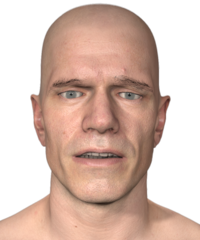}} & \hspace{\hgap}
\CenterImage{\includegraphics[width=\rowlabelwidth]{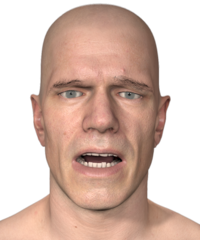}} & \hspace{\hgap}
\CenterImage{\includegraphics[width=\rowlabelwidth]{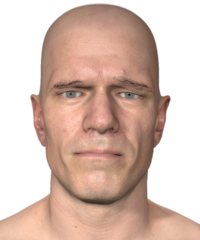}} & \hspace{\hgap}
\CenterImage{\includegraphics[width=\rowlabelwidth]{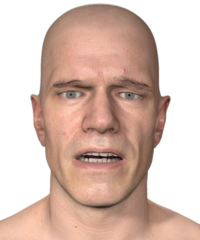}} \\
\end{tabular}\\[0.2em]

\textcolor{gray!60}{\rule{\textwidth}{1pt}}\\[0.2em]

\begin{flushleft}
(c) Intensity Control
\end{flushleft}
\vspace{-0.5em}
\begin{tabular}{m{3.1cm} c c c c c c c c }
\parbox{2.9cm}{ \centering \promptText{%
   ``a person looks angry''\\
   \vspace{-1em} $\xdownarrow{}$\\
  ``a person looks extremely angry''
}}&
\CenterImage{\includegraphics[width=\rowlabelwidth]{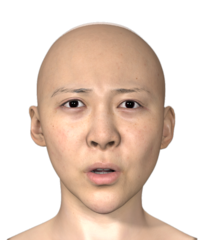}} & \hspace{\hgap}
\CenterImage{\includegraphics[width=\rowlabelwidth]{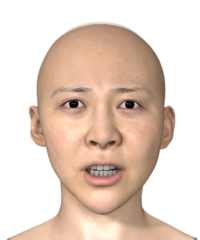}} & \hspace{\hgap}
\CenterImage{\includegraphics[width=\rowlabelwidth]{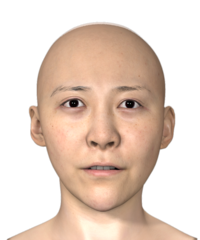}} & \hspace{\hgap}
\CenterImage{\includegraphics[width=\rowlabelwidth]{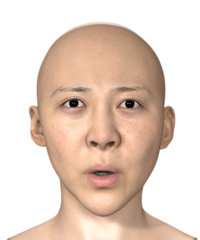}} & \hspace{\hgap}
\CenterImage{\includegraphics[width=\rowlabelwidth]{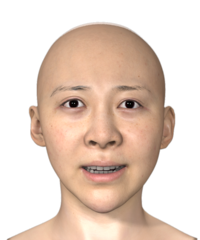}} & \hspace{\hgap}
\CenterImage{\includegraphics[width=\rowlabelwidth]{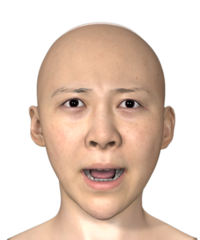}} & \hspace{\hgap}
\CenterImage{\includegraphics[width=\rowlabelwidth]{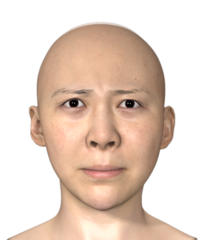}} & \hspace{\hgap}
\CenterImage{\includegraphics[width=\rowlabelwidth]{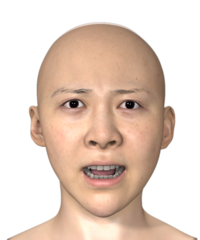}} \\
& \emph{\footnotesize \strut Anger } &

\multicolumn{6}{c}{%
\raisebox{0.05\rowlabelwidth}{ 
\tikz[baseline]{  
  \draw[->, thick, color=nvidiagreen, line width=2pt, >=stealth]
    (0,0) -- (5.0\rowlabelwidth,0);
}
}} & \hspace{\hgap}
\emph{\footnotesize \strut ANGER} \\
\end{tabular}

\textcolor{gray!60}{\rule{\textwidth}{1pt}}\\[0.2em]

\begin{flushleft}
(d) Emotion Transitions
\end{flushleft}
\begin{tabular}{m{3.1cm} c c c c c c c c }
\parbox{2.9cm}{ \centering \promptText{%
\shortstack[c]{``a person looks joyful''\\$\xdownarrow{}$\\``a person looks angry''}}}&
\CenterImage{\includegraphics[width=\rowlabelwidth]{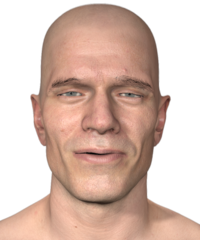}} & \hspace{\hgap}
\CenterImage{\includegraphics[width=\rowlabelwidth]{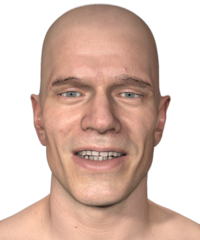}} & \hspace{\hgap}
\CenterImage{\includegraphics[width=\rowlabelwidth]{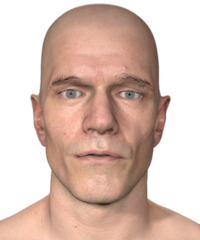}} & \hspace{\hgap}
\CenterImage{\includegraphics[width=\rowlabelwidth]{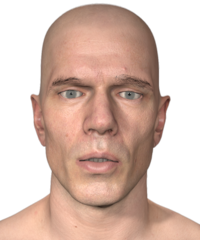}} & \hspace{\hgap}
\CenterImage{\includegraphics[width=\rowlabelwidth]{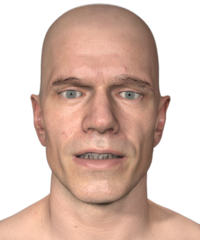}} & \hspace{\hgap}
\CenterImage{\includegraphics[width=\rowlabelwidth]{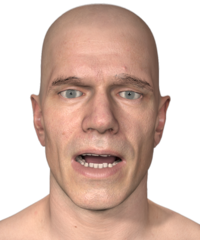}} & \hspace{\hgap}
\CenterImage{\includegraphics[width=\rowlabelwidth]{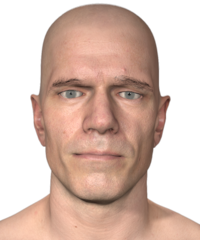}} & \hspace{\hgap}
\CenterImage{\includegraphics[width=\rowlabelwidth]{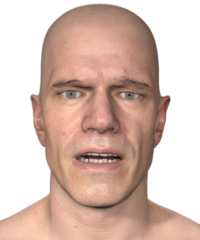}} \\
& \emph{\footnotesize \strut Joy } &
\multicolumn{6}{c}{%
\raisebox{0.05\rowlabelwidth}{ 
\tikz[baseline]{
  \draw[->, thick, color=nvidiagreen, line width=2pt, >=stealth]
    (0,0) -- (5.0\rowlabelwidth,0);
}
}
} & \hspace{\hgap}
\emph{\footnotesize \strut Anger} \\
\end{tabular}

\caption{The results of text-driven emotion and facial movement control. (a) Emotion control, including the unseen emotion `lonely'. (b) Facial movement control with neutral emotion. (c) Intensity control from low to high levels of anger. (d) Emotion transitions from joy to anger through interpolation between text embeddings.}
\label{fig:results_text_cond}
\end{figure}
\setlength{\rowlabelwidth}{0.12\textwidth}
\setlength{\hgap}{-0.9em} 

\begin{figure}[H]
\centering
\setlength{\tabcolsep}{0pt}     
\renewcommand{\arraystretch}{0} 

\begin{tabular}{c c c c c c c c c c}
& \emph{\nvGreen{w}ith} & \hspace{\hgap} 
\emph{t\nvGreen{e}nure} & \hspace{\hgap} 
\emph{S\nvGreen{u}zie} & \hspace{\hgap} 
\emph{\nvGreen{a}ll} & \hspace{\hgap} 
\emph{l\nvGreen{ei}sure} & \hspace{\hgap} 
\emph{\nvGreen{f}or} & \hspace{\hgap} 
\emph{y\nvGreen{a}tching}& \hspace{\hgap} 
\emph{\nvGreen{b}ut} & \hspace{\hgap} 
\emph{g\nvGreen{oo}d}\\[0.3em]
\rotatebox{90}{\makebox[0pt][l]{\emph{\small v3.0 \footnotesize (James id)}}} \hspace{0.1em} & 
\includegraphics[width=\rowlabelwidth]{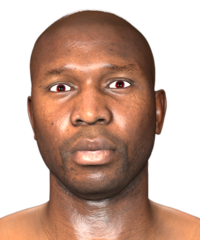} & \hspace{\hgap} 
\includegraphics[width=\rowlabelwidth]{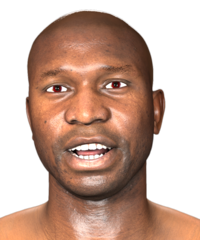} & \hspace{\hgap} 
\includegraphics[width=\rowlabelwidth]{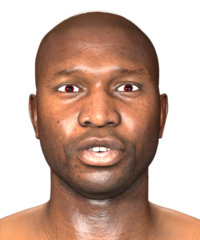} & \hspace{\hgap} 
\includegraphics[width=\rowlabelwidth]{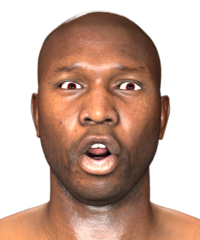} & \hspace{\hgap} 
\includegraphics[width=\rowlabelwidth]{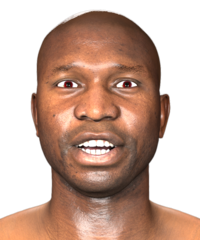} & \hspace{\hgap} 
\includegraphics[width=\rowlabelwidth]{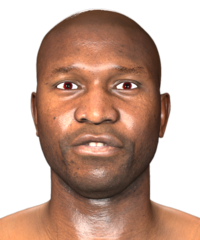} & \hspace{\hgap} 
\includegraphics[width=\rowlabelwidth]{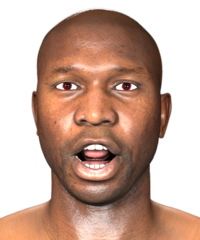} & \hspace{\hgap} 
\includegraphics[width=\rowlabelwidth]{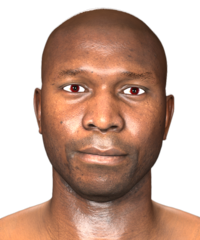} & \hspace{\hgap} 
\includegraphics[width=\rowlabelwidth]{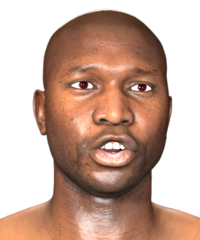}  \\[0.5em]
\rotatebox{90}{\makebox[0pt][l]{\emph{\hspace{0.1em} \small Head Rotations}}} \hspace{0.1em} & 
\multicolumn{9}{c}{%
\includegraphics[width=8.2\rowlabelwidth]{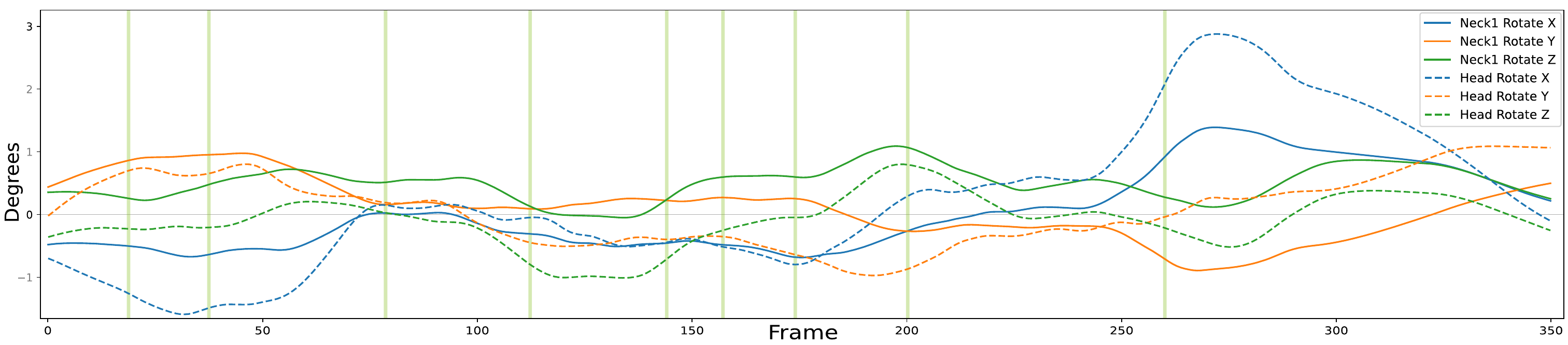}
}
\end{tabular}
\vspace{-0.5em}
\caption{\emph{Head motion results.} \emph{Top:} Example frames showing synthesized head and neck rotations. Root motion excluded for visual clarity. \emph{Bottom:} Head and neck rotation curves illustrating smooth, natural dynamics over time. Green vertical lines highlight the frames depicted above.}
\label{fig:result_head_motion}
\end{figure}

\subsection{Head Motion Generation}
To produce more natural full-head animation, we extend the diffusion model to generate head motion animation together with facial animation by including the head motion as an additional generation target in the diffusion model. The diffusion model is modified to include additional input projection and output decoder layers for head motion. The head motion is parameterized by a matrix $\mH \in \R^{N \times 9}$, where $N$ is the number of frames, and each frame consists of 6 degrees of freedom for the 2 head joints (each with 3 Euler rotation angles) and 3 degrees of freedom for root translation. To reduce the head motion jitter, we also include head motion acceleration loss in the training loss:
\begin{equation}
\gL_{\mathrm{accel}} =  \left\|\left( \hat{\mH}_{n+1} - 2\hat{\mH}_{n} + \hat{\mH}_{n-1} \right)\right\|_F^2.
\end{equation}

Since we only have a few characters with head motion data, to handle missing head motion data and train a multi-id model with head motion for all identities, we use a special training approach: 
\begin{itemize}
    \item For training samples from identities without ground truth head motion, we exclude all head-related loss terms.
    \item For training samples from identities with head motion ground truth, with a certain probability, we substitute the input identity vector $\vi$ with a different identity vector $\vi'$ without head motion data, and only apply head-related loss.
\end{itemize}

In addition, to generate natural idle motions during periods of silence, we include some additional idle motion data in the training data, where the actor is instructed to perform idle motions. \cref{fig:result_head_motion} shows example animations generated by the model.

\subsection{Audio-driven Facial Rig Parameters}
\label{sec:a2frig}

The presented \audiotoface pipeline first infers dense facial geometry, followed by a blendshape solving step to estimate rig weights. However, many applications only require the final rig parameters for animation or retargeting, making the intermediate geometry superfluous. To enable tighter integration with production rigs using a simpler network, we experimented with a unified system, which directly predicts facial rig parameters from audio input.

\subsubsection{Dataset Construction for Rig-driven Supervision}

\textbf{Identity-Agnostic Rig Design.}
To improve generalization across characters, we train on a single, generic facial rig rather than personalized rigs tied to different \audiotoface identities. This generic identity is constructed by blending multiple existing \audiotoface characters to produce a neutral base. Additionally, we keep the rig's shape expressions close to the canonical ARKit blendshape set, which makes the resulting model easier to retarget across different characters and platforms.

\textbf{Weight Extraction from 4D Sequences.}
To build training data, we convert 4D captured expressions from~\cref{subsec:capture} into blendshape weights using the blendshape solve method in~\cref{sec::bs_solve}. For more complex rigs that incorporate corrective or composite shapes, such as MetaHumans~\citep{epic2021metahuman}, a quartic solver~\citep{rackovic2024quartic} can be used for more accurate results.

\textbf{Data Post-Processing.}
The raw weights obtained from solver inference are often not sufficient for direct training due to semantic mismatches, inconsistencies in control behavior, or temporal instability. We apply a series of post-processing steps to ensure that the extracted weights better reflect the intent of the original performance and are robust enough to serve as training targets:

\begin{itemize}
    \item \textbf{Tooth Exposure and mouth opening:} Adjust upper and lower teeth visibility and jaw position to match the observed occlusion and mouth opening, correcting cases where the solver under/over-emphasizes jaw or lip motion.
    
    \item \textbf{Eye White Exposure:} Match sclera visibility to better calibrate the intensity of eye expressions. This is especially important for emotion-driven sequences such as anger, fear, or surprise, where eye openness is a strong semantic signal.

    \item \textbf{Viseme Refinement:} Emphasize the articulation of key phoneme-related shapes such as \texttt{M/B/P} (lip closure, press, and rounding) and \texttt{F/V} (lip-to-teeth contact and lip roll), which are often critical for speech intelligibility.

    \item \textbf{Temporal Filtering:} Reduce jitter and high-frequency noise in the solved weight sequences, ensuring temporal consistency and smoother animation.
\end{itemize}

\subsubsection{Direct Inference of Rig Parameters}
\label{sec:a2frig-architecture}

The architecture of the rig-parameters network is similar to the \audiotoface diffusion model. The network follows a diffusion-based approach but with a simplified encoder-decoder structure. Instead of using input projection layers for each facial component, the model directly concatenates the noisy rig parameters with other input features (audio feature, emotion embedding, identity embedding, and timestep embedding) and feeds them into a GRU network. The GRU output is then decoded to produce denoised rig parameters using a linear layer. In addition, a mixture of experts (MoE) layer~\citep{jiang2024mixtral} is used inside the Hubert encoder to allow the model to handle id-specific behavior and leads to less coupling between facial animation across identities. The MoE layer replaces the intermediate dense layer in the Hubert feed-forward network with $N$ expert networks, where $N$ corresponds to the number of identities. Each expert $E_i$ applies a linear transformation:
\begin{equation}
E_i(\vh) = \mM_i \vh + \vb_i,
\end{equation}
where $\mM_i \in \R^{d_{\text{intermediate}} \times d_{\text{hidden}}}$ and $\vb_i \in \R^{d_{\text{intermediate}}}$ are the expert-specific parameters. The MoE output is computed using identity-based routing using identity vector $\vi$:
\begin{equation}
\vh_{\text{MoE}} = \sum_{i=1}^N \vi_i \cdot E_i(\vh).
\end{equation}
The model is trained using a similar denoising objective as the \audiotoface diffusion model.

\setlength{\rowlabelwidth}{0.1375\textwidth}

\begin{figure}[t]
\centering
\setlength{\tabcolsep}{0pt}     
\renewcommand{\arraystretch}{0} 

\setlength{\hgap}{-2.1em} 

\begin{tabular}[t]{c c}
\begin{tabular}[t]{c c c c c c c c}
     & \emph{t\nvGreen{e}nure} & \hspace{\hgap} \emph{S\nvGreen{u}zie} & \hspace{\hgap} \emph{\nvGreen{a}ll} & \hspace{\hgap} \emph{l\nvGreen{ei}sure} & \hspace{\hgap} \emph{\nvGreen{f}or} & \hspace{\hgap} \emph{y\nvGreen{a}tching}  & \hspace{\hgap} \emph{\nvGreen{b}ut} \\[0.3em]
        \rotatebox{90}{\makebox[0pt][l]{\emph{\small Metahuman Rig}}} \hspace{0.1em} & 
        \includegraphics[width=\rowlabelwidth]{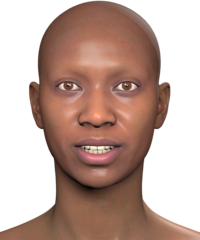} & \hspace{\hgap}
        \includegraphics[width=\rowlabelwidth]{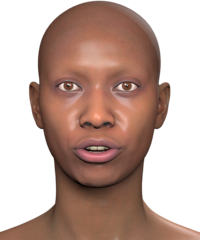} & \hspace{\hgap}
        \includegraphics[width=\rowlabelwidth]{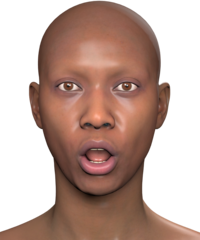} & \hspace{\hgap}
        \includegraphics[width=\rowlabelwidth]{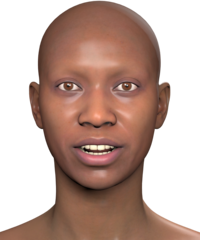} & \hspace{\hgap}
        \includegraphics[width=\rowlabelwidth]{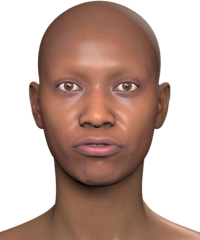} & \hspace{\hgap}
        \includegraphics[width=\rowlabelwidth]{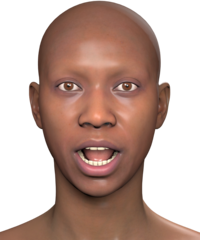} & \hspace{\hgap}
        \includegraphics[width=\rowlabelwidth]{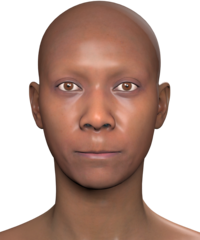} \\
\end{tabular} 
&
\hspace{-0.3em} \raisebox{-9.0em}{\includegraphics[width=1.75\rowlabelwidth]{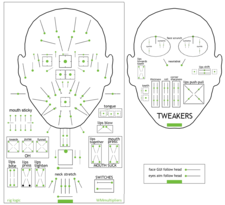}}
\end{tabular}
\vspace{-0.75em}
\caption{\emph{left}: Results of the rig-parameters network applied on an example Metahuman rig for the same audio input as in \cref{fig:results_bs_solve}. \emph{right}: The network output is applied to the control rig of the character for the frame \emph{\nvGreen{a}ll}.}
\label{fig:result_A2FR}
\end{figure}

\subsection{Jaw-driven Blendshape Solver}
\label{sec::jaw_constraint}

\begin{figure}[!t]
    \centering    
    \begin{tabular}{c c c c c}
        \emph{\small A2F Inference} & \hspace{-3.6em} \emph{\small Base Solver} & \hspace{-3.6em} \emph{\small Jaw-driven Solver} &  \emph{\small \texttt{jawOpen} Activation} \\
        \includegraphics[height=0.25\textwidth]{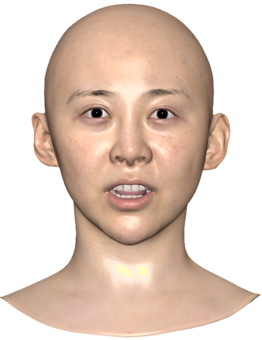}      & \hspace{-3.6em}
        \includegraphics[height=0.25\textwidth]{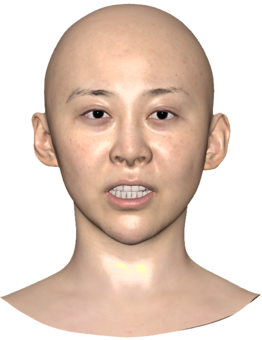}     & \hspace{-3.6em}
        \includegraphics[height=0.25\textwidth]{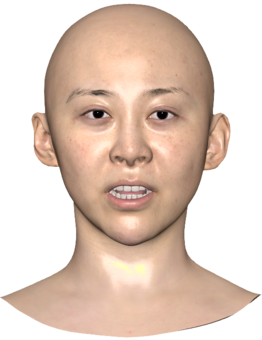} &        
        \vspace{1em} \includegraphics[height=0.24\textwidth]{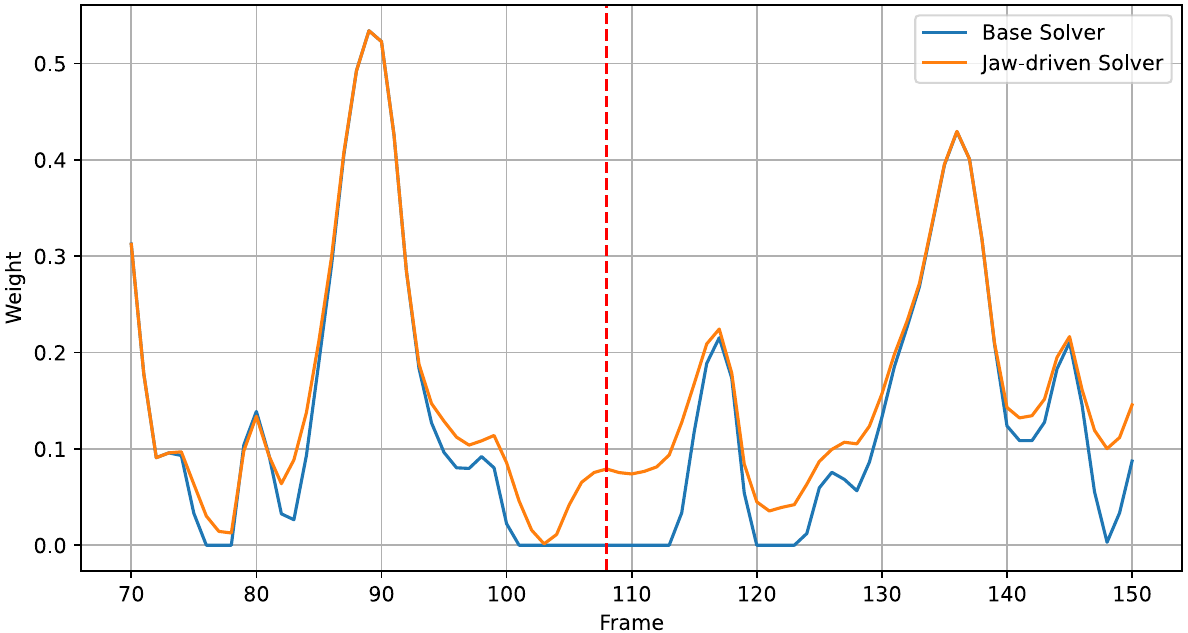} \\
    \end{tabular}
    \vspace{-1.5em}
    \caption{
        \emph{Solver comparison.} \emph{Left:} A2F inference, base solver, and jaw-driven solver results. \emph{Right:} ARKit \texttt{jawOpen} weights over time for both solvers; the red dashed line indicates the frame shown on the left.
    }
    \label{fig:jaw_driven_solver}
\end{figure}

Our blendshape solver estimates facial expression weights from surface geometry alone. However, jaw motion is often hard to disambiguate from skin deformation, especially when the blendshape model cannot fully represent the lower face’s articulation. This can result in solutions where the facial surface fits geometrically well, but the jaw position is physically inaccurate, an issue that is particularly problematic in speech-driven expressions. For example, the solver may miss the open lips and closed jaw characteristic of \textit{EE} visemes. To address this, we introduce a soft constraint that integrates the jaw displacement inferred by \audiotoface into the optimization. This jaw-driven approach improves anatomical correctness by better matching target jaw articulation while preserving skin surface fidelity, capturing secondary jaw-related motions often missed by the base solver (\cref{fig:jaw_driven_solver}).
 
\textbf{Constraint Formulation.}
We encode jaw motion as a linear blend of jaw-specific blendshape deltas, captured in a matrix $\mathbf{C} \in \R^{3 \times N}$, where each column represents the 3D displacement vector of a jaw-related shape at a reference vertex (e.g., the jaw center). Given an observed jaw delta vector $\vd_{\text{jaw}} \in \R^{3}$, we penalize the deviation between this signal and the reconstructed motion:

\begin{equation}
R_{\text{jaw}}(\vw) = \lambda_{\text{jaw}} \left\| \mathbf{C} \vw - \vd_{\text{jaw}} \right\|_2^2.
\end{equation}

\textbf{Dynamic Scaling.}
While jaw position is important for anatomical plausibility, enforcing it too rigidly can harm overall quality. Because blendshape models have limited expressiveness, strictly matching the jaw position can force unnatural shape combinations that fail to fit the skin surface accurately. To balance anatomical correctness with visual fidelity, we treat jaw motion as a soft constraint by dynamically activating its influence when correct jaw alignment is critical and relaxing it when the jaw is clearly articulated or less visually exposed. 

\begin{equation}
\lambda_{\text{jaw}} = \lambda_0 \cdot \exp\left(-\frac{\|\vd_{\text{jaw}}\|^2}{2\sigma^2} \right) \cdot \left(1 - \exp\left(-\frac{\delta_{\text{lip}}^2}{2\sigma^2} \right)\right).
\end{equation}

 The constraint weight $\lambda_{\text{jaw}}$ is adjusted dynamically based on the target jaw displacement $|\vd_{\text{jaw}}|$ and the central vertical distance between lips $\delta_{\text{lip}}$. It decreases with large jaw displacements, allowing for more freedom when jaw motion is clearly expressed, and increases when the jaw is relatively static to encourage correct positioning in closed-mouth cases. At the same time, the weight increases with lip separation, focusing the constraint when the jaw is visually exposed and reducing it otherwise. 
 The sensitivity parameter $\sigma$ is tuned per subject to match typical jaw displacement magnitudes in the training data.

\section{Conclusion}
\label{sec::conclusion}

In this work, we introduce \audiotoface, an advanced audio-driven facial animation system for digital avatars. Leveraging a high-quality 4D capture dataset, we train both a regression-based network and a diffusion-based network. The network outputs are subsequently post-processed and converted into rig parameters. Our \audiotoface system generates realistic lip sync animations from any speech input, irrespective of the voice or language, and operates in real-time to facilitate natural interactions with interactive avatars as well as offline facial animation authoring. In conjunction with Audio2Emotion, digital avatars can express various emotions without the need for manual emotion specification.

We also explore several innovative concepts, including text-conditioned emotion, head motion animation, the direct generation of rig parameters from audio, and jaw-driven blendshape solve. To support digital avatar creators and industry professionals, we have made our networks, SDK, and training framework open-weight and open-source, thereby contributing to the democratization of digital human technology.


\clearpage
\appendix
\section{Audio2Emotion}
\label{sec::audio2emotion}
Animating a face convincingly requires capturing the subtle emotional shifts that accompany speech. Although manual adjustment of emotional expressions using predefined emotion sliders in \audiotoface is possible, this approach is labor-intensive and insufficient to track dynamic emotional nuances within spoken sentences. To automate this, we introduce Audio2Emotion, a neural network designed to extract a continuous emotional signal directly from speech audio. This emotional signal is then seamlessly integrated with \audiotoface, producing natural and expressive facial animations in real time without manual intervention.

\subsection{Design}

At the core of Audio2Emotion lies a six-class speech-emotion classifier. Due to the absence of publicly available datasets with continuous emotion annotations, we rely instead on a classifier trained on discrete emotion labels. The model classifies audio into six basic emotions: anger, disgust, fear, joy, neutral, and sadness, selected based on the availability of labeled datasets.

To derive a continuous emotional sequence from the utterance-level classifier, Audio2Emotion employs a sliding-window approach. The input audio is chopped into a sequence of overlapping windows, each of which is analyzed by the network. The algorithm is naturally controlled by two user-tunable hyperparameters, controlling the trade-off between reliability and temporal detail:
\begin{itemize}
    \item Window size (default $\approx$ 1.9~s). A larger window yields more reliable class estimates but compresses variation: in the limit case, if the window spans the whole utterance, only one emotion is produced. The default value was chosen empirically as the shortest context that still yields reliable predictions.
    \item Stride (default 0.5~s). A smaller stride samples the clip more often, increasing temporal detail but may introduce jitter in the timeline; a larger stride has the opposite effect.
\end{itemize}

To suppress over-confident spikes and encourage smoother transitions, we apply a second, finer sweep within every window. Each 1.9~s slice is subdivided into overlapping 0.625~s sub-windows (stride $\approx$ 0.31~s); the six class probabilities are averaged across these sub-windows before being sent to \audiotoface. This “double sliding window” dampens the tendency of the network to over-estimate confidence for a single class and produces more realistic blends of emotions.

If the full audio is available (offline use) we simply linearly interpolate the resulting keyframes. When audio is streamed (online use), we replace interpolation with a lightweight exponential smoothing, yielding a continuous, low-latency emotion track that drives natural upper-face motion in real time.

\subsection{Network Architecture and Training Details} Audio2Emotion is implemented as a six-class speech-emotion classifier fine-tuned from \textit{facebook/wav2vec2-large-lv60} ($\approx$ 315 M parameters). The CNN feature extractor remains frozen; the 24-layer Transformer and a classifier head are trained end-to-end. Input waveforms are resampled to 16 kHz, truncated up to 10 seconds, and Z-normalized. Optimization uses AdamW with a learning rate schedule that decreases linearly from $5\times 10^{-4}$ to $5 \times 10^{-5}$ over 20 epochs. The batch size is equal to 16.

\subsection{Data}

Public resources for speech-emotion recognition are still modest: most contain only a few thousand clips, and many cannot be used in commercial products.
To assemble a workable training set, we therefore combined all corpora that met three practical criteria:
\begin{enumerate}
    \item The emotion labels match the six classes used in Audio2Emotion (anger, disgust, fear, joy, neutral, sadness).
    \item The language is European, as cultural differences in emotional expression and vocal patterns can significantly impact recognition accuracy.
    \item The license permits commercial use, or a separate paid license is available (e.g., RAVDESS).
\end{enumerate}

\cref{tab:a2e-datasets} lists all the corpora used for training and evaluation.

For training, we use five publicly available datasets (RAVDESS, CREMA-D, JL Corpus, EMO-DB, Emozionalmente), along with two additional datasets. The first is A2E-OpenAI-TTS, an emotional speech dataset generated using OpenAI GPT-4o’s advanced text-to-speech capabilities. We designed prompts to elicit a wide range of emotions and found that GPT-4o can produce surprisingly realistic and expressive audio, making this resource valuable for both data diversity and quantity.
The second is a \textit{Private-Train} dataset collected internally. Due to privacy constraints, these datasets will not be released.

For evaluation, we recorded our own \textit{Private-Test} dataset of 27 speakers, none of whom appears in the training data. This set is used exclusively for internal benchmarking and is not released due to privacy constraints.
Furthermore, we evaluated the model on three publicly available datasets (TESS, SAVEE, and IEMOCAP) for a broader comparison. For IEMOCAP, we follow the standard four-class protocol: anger; happy (including excitement); neutral; and sad—discarding other categories for consistency with related work.

\begin{table}[t]
\centering
\caption{Speech–emotion corpora used for training and evaluation.}
\label{tab:a2e-datasets}
\resizebox{\textwidth}{!}{%
\begin{tabular}{lccc}
\toprule
\multicolumn{4}{c}{\textbf{Training corpora}} \\ \cmidrule(lr){1-4}
Dataset & \# Utterances & \# Speakers & Emotion Classes \\ \midrule
RAVDESS \citep{a2e_radvess}        & 1056 & 24 &  Anger, Disgust, Fear, Joy, Neutral, Sad \\
CREMA-D \citep{a2e_cremad} & 7441 & 91 &   Anger, Disgust, Fear, Joy, Neutral, Sad \\
JL Corpus \citep{a2e_jl}             & 1200 & 4 &  Anger, Joy, Neutral, Sad \\
EMO-DB \citep{a2e_emodb}             & 454 & 10 &  Anger, Disgust, Fear, Joy, Neutral, Sad \\
Emozionalmente \citep{a2e_emoz}             & 1500 & 303 &  Anger, Disgust, Fear, Joy, Neutral, Sad \\
A2E-OpenAI-TTS             & 9954 & 9 &  Anger, Disgust, Fear, Joy, Neutral, Sad \\
\textit{Private-Train} & 8424 & 2 &  Anger, Disgust, Fear, Joy, Neutral, Sad \\ 
\addlinespace
\multicolumn{4}{c}{\textbf{Evaluation corpora}} \\ \cmidrule(lr){1-4}
TESS \citep{a2e_tess}     & 2400 & 2 &  Anger, Disgust, Fear, Joy, Neutral, Sad \\
SAVEE \citep{a2e_savee}     & 420 & 4 &  Anger, Disgust, Fear, Joy, Neutral, Sad \\
IEMOCAP \citep{a2e_iemocap}     & 5531 & 10 &  Anger, Happy (+Excitement), Neutral, Sad \\
\textit{Private-Test}  & 1350 & 27 &  Anger, Disgust, Fear, Joy, Neutral, Sad \\ \bottomrule
\end{tabular}

}
\end{table}

\subsection{Speaker-specific Personalization}

We observed that increasing the number of unique speakers in the training data improves generalization much more than adding additional samples per speaker. However, even with this strategy, speaker variation in real-world use still leads to inconsistent emotion recognition performance. To address this, we developed two practical methods for speaker-specific personalization.

\textbf{Fine-tuning on Speaker-specific Data.}
In this approach, a small set of labeled utterances is collected from the target speaker (typically 1 to 3 samples per emotion). The generic Audio2Emotion model is then slightly fine-tuned on this personalized dataset, which takes a few seconds to a few minutes on a modern GPU. This method significantly improves the classification accuracy for the specific speaker, while the model retains reasonable performance on other voices, indicating that it adapts without catastrophic forgetting.

\textbf{Personalization Using a Neutral Sample.}
To reduce the burden on the user, we propose a lighter personalization strategy that leverages a single neutral utterance from the target speaker. For each emotional utterance to be classified (the target sample), we concatenate it with a neutral sample from the same speaker, separated by two seconds of silence. The model is then trained to classify this concatenated input according to the emotion label of the target sample. At inference time, when the label is unknown, we apply the same concatenation using the available neutral sample, and use the model’s prediction for the target segment as the emotion output. This method requires only a single neutral recording from the user, yet still provides a noticeable improvement in emotion recognition accuracy with minimal user effort.

Empirically, the performance of these approaches can be ranked as follows:$$
\text{generic model} \leq \text{neutral sample-based personalization} \leq \text{fine-tuning on speaker data.}$$
Thus, the user can choose the appropriate trade-off between the personalization effort and the accuracy of emotion recognition, depending on the application requirements.

\subsection{Experiments and Evaluation}

\begin{figure}[htbp]
  \centering
  
  \begin{subfigure}[t]{0.48\textwidth}
    \includegraphics[width=\linewidth]{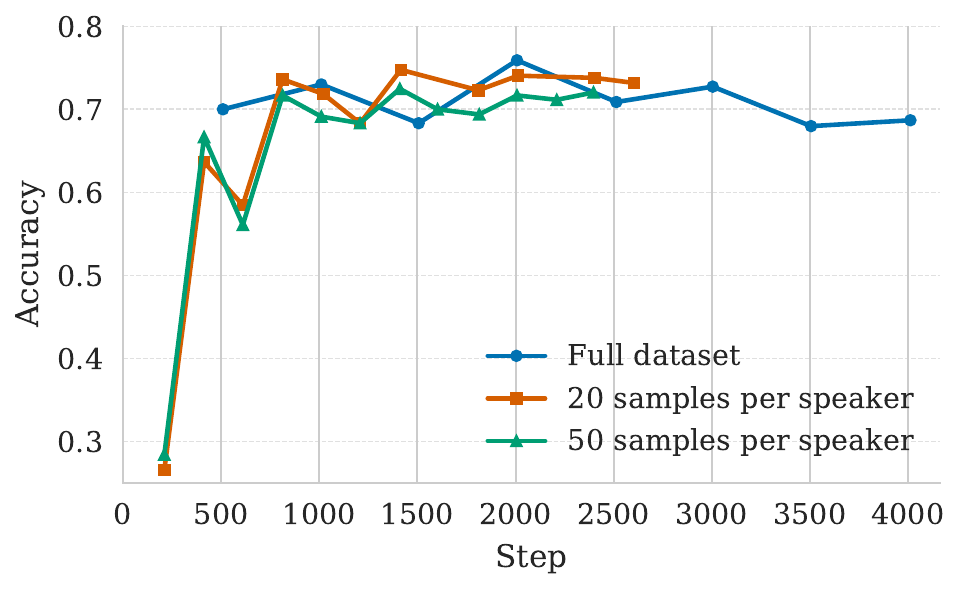}  
    \caption{Model accuracy over course of training for different training datasets: full dataset or reduced to 20/50 samples per speaker. The models exhibit identical behavior, showing that adding more data per speaker adds no noticeable value.}
    \label{fig:a2e_data_research_a}
  \end{subfigure}
  \hfill                   
  \begin{subfigure}[t]{0.48\textwidth}
    \includegraphics[width=\linewidth]{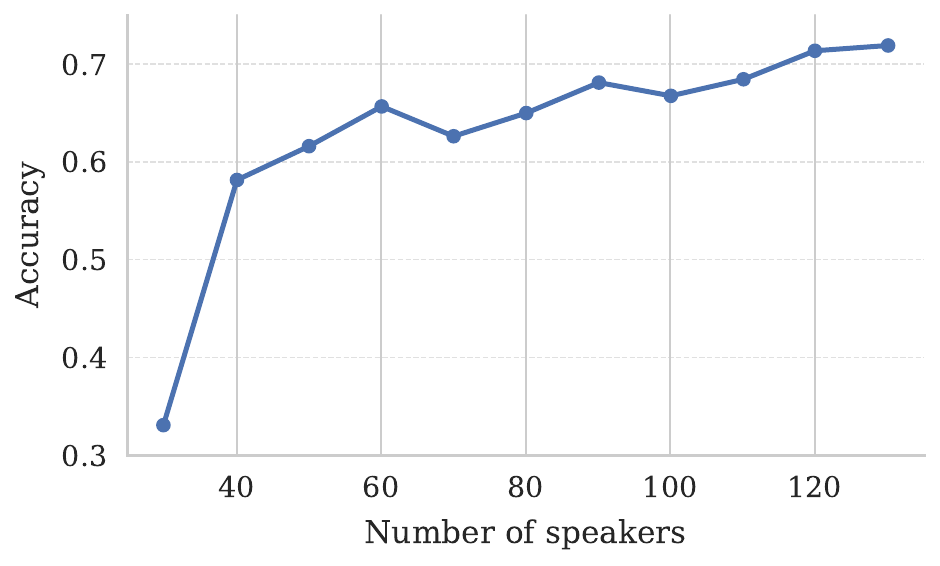}
    \caption{Median accuracy of models depending on the amount of speakers in the training dataset. The plot shows that the amount of speakers in training data is a strong contributor to resulting accuracy.}
    \label{fig:a2e_data_research_b}
  \end{subfigure}
  
  \caption{Experiments with training data.}
  \label{fig:a2e_data_research}
\end{figure}

\begin{figure}[h]
    \centering
    \includegraphics[width=0.6\textwidth]{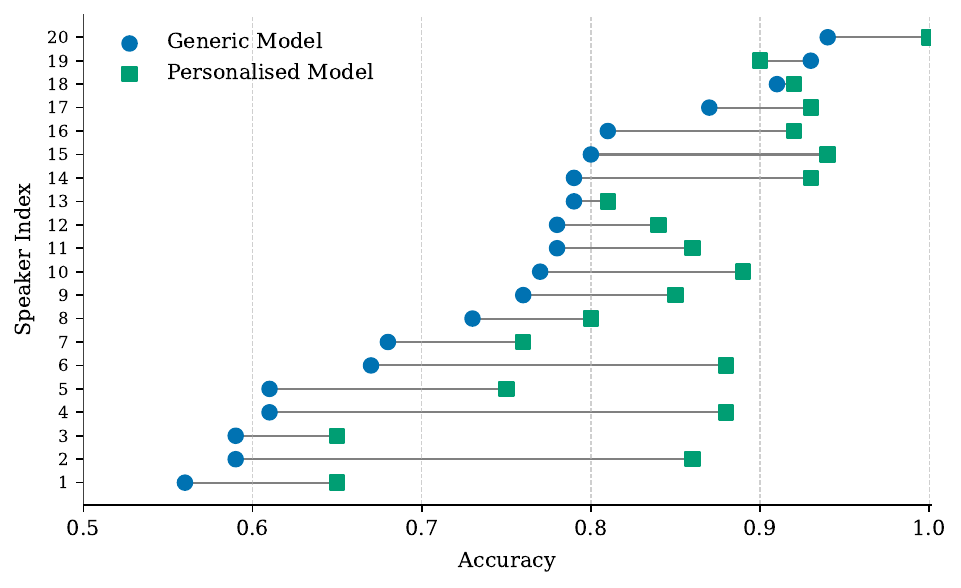}
    \caption{Accuracy of the generic versus personalised classifiers for each of the 20 speakers after fine-tuning on 12 speaker-specific clips (2 per emotion). The single decline (speaker \#19) is attributed to poorly selected random clips in that speaker’s personalization dataset.}
    \label{fig:a2e_pers_gains}
\end{figure}

\textbf{Evaluation Methodology.} Although our end goal is to generate facial animations that look natural to human observers, the only tractable metric we can report is clip-level classification accuracy: no agreed-upon quantitative measure of animation quality exists, and emotion timelines are intrinsically ambiguous. We treat accuracy as a proxy metric that is correlated with animation realism: higher scores generally yield smoother, more convincing faces, but once accuracy approaches its ceiling of $\approx$ 70\%, the correlation flattens, and reliable assessment requires manual inspection.

One of our early findings was that including the same speakers in both training and test splits can inflate accuracy, as the network begins to rely on speaker-specific cues rather than true emotion generalization. To prevent this, all of our benchmarks are strictly speaker-independent: no speaker in any test set appears in the training data.

\textbf{Training Data Experiments.} \cref{fig:a2e_data_research_a} plots clip-level accuracy over training when we progressively cap the number of utterances per speaker (full data, 50 clips, and 20 clips). All three curves oscillate in the same narrow band, indicating that once a speaker contributes $\approx$ 20 examples, adding more utterances provides no measurable gain.
\cref{fig:a2e_data_research_b} shows the complementary experiment: we fix 20 clips per speaker but vary the number of speakers in the training set. The median accuracy grows steadily with speaker count, confirming that speaker diversity is the dominant driver of generalization.

\textbf{Checkpoints.} Our open-source package provides two pretrained ONNX models:

\begin{itemize}
\item \textbf{A2E-v2.2 (stable).} Trained on full-length audio segments with random-crop augmentation. It achieves slightly higher classification accuracy on average, but can overestimate confidence for some incorrect predictions, which may degrade downstream animation quality.
\item \textbf{A2E-v3.0 (research preview).} Trained on fixed 0.625~s crops and designed for double sliding-window inference to mitigate confidence overestimation. This inference procedure is embedded in the ONNX graph, so no additional sub-segmentation of the input audio is required.
\end{itemize}

For completeness, we also report metrics for \textbf{A2E-Personalized}, a lightweight model that conditions on a single neutral utterance from the target speaker (see the Personalization subsection). This checkpoint is not released because it is not yet integrated into the \audiotoface API, and our license prohibits standalone use of Audio2Emotion.

\textbf{Personalization Experiments.} \cref{fig:a2e_pers_gains} reports results for the subset of 20 test speakers who had a sufficient amount of data for the personalization procedure. For every speaker, we sample just 12 clips—two for each of the six emotions—and fine-tune the generic model for a few seconds on a single GPU. The remaining clips of that speaker form the evaluation set. Personalization improves accuracy for 19 of the 20 speakers with an average accuracy gain of 10\%; the single drop (speaker \#18) is explained by an unlucky random draw of training clips.

\textbf{Evaluation Results.} \cref{tab:a2e-evaluation} reports clip-level accuracy for the three checkpoints. \textbf{A2E-Personalized} achieves the best overall scores. Among generic models, \textbf{v2.2} is stronger on public benchmarks (IEMOCAP-4/TESS/SAVEE), while \textbf{v3.0} leads on \emph{Private-Test} and yields better-calibrated frame-wise predictions for animation.

\textbf{Human-feedback fine-tuning.} Complementary work \citep{a2e_rlhf} shows that Direct Preference Optimization \citep{a2e_dpo} on crowdsourced pairwise ratings can further refine Audio2Emotion for animation quality without explicit emotion labels. We refer interested readers to that paper for details and results.

\begin{table}[t]
\centering
\caption{Emotion classification accuracy on evaluation datasets.}
\label{tab:a2e-evaluation}
\begin{tabular}{lcccc}
\toprule
Checkpoint & Private-Test & IEMOCAP-4 & TESS & SAVEE \\ \midrule
Audio2Emotion-v2.2 & 0.64 & 0.46 & 0.86 & 0.66 \\
Audio2Emotion-v3.0 & 0.69 & 0.41 & 0.71 & 0.63 \\
Audio2Emotion-Personalized & 0.79 & 0.44 & 0.92 & 0.74 \\ \bottomrule
\end{tabular}
\end{table}

\clearpage
\section{Contributors and Acknowledgments}
\label{sec::contributors}

\subsection{Core Contributors}
Chaeyeon Chung,
Ilya Fedorov, 
Michael Huang,
Aleksey Karmanov,
Dmitry Korobchenko,
Roger Ribera, 
Yeongho Seol

The core contributors (listed in alphabetical order) were involved in various aspects of \audiotoface and Audio2Emotion research work, including data curation and processing, designing network architectures, training and testing networks, improving the output quality, experimenting with new features, organizing and writing this paper.

\subsection{Acknowledgments}
We'd like to thank Martin Bisson, Joey Lai, David Minor, Simon Ouellet, Sehwi Park for their hard engineering efforts in integrating \audiotoface technology into the SDK, Maya ACE, and downstream applications in an optimal way. We appreciate Cyrus Hogg, Edy Lim, Simon Yuen for their project and quality guidance. We thank Ronan Browne, Miguel Guerrero, Ehsan Hassani, Marco Di Lucca, Hans Yang for their contribution to data and asset generation. Finally, we are grateful to Sherry Faramarz, Anna Minx, Will Telford for project management and release support.

\clearpage
\setcitestyle{numbers}
\bibliographystyle{plainnat}
\bibliography{main}

\end{document}